%% file: paper_followup.tex
\documentclass[a4paper,11pt]{article}
\usepackage{jinstpub}

\usepackage[sicmds,freestanding]{hepunits}
\usepackage{booktabs}
\usepackage{subcaption}
\usepackage{placeins}
\usepackage{tikz}
\usetikzlibrary{arrows.meta, positioning, shapes.geometric, fit,positioning,decorations.markings}
\usetikzlibrary{calc}

\title{\boldmath Hybrid neural denoising for resource-efficient near- and sub-threshold radio triggering of extensive air showers}

\author[b,1]{Qader Dorosti\note{Corresponding author.}}
\author[a]{\unskip,\,Alperen Aksoy}
\author[a]{\unskip,\,Ilja Bekman}
\author[b]{\unskip,\,Markus Cristinziani}
\author[b]{\unskip,\,Eric-Teunis de Boone}
\author[a,b]{\unskip,\,Vesselin Dimitrov}
\author[a]{\unskip,\,Chimezie Eguzo}
\author[c]{\unskip,\,Stefan Heidbrink}
\author[a,d]{\unskip,\,Stefan van Waasen}
\author[a]{\unskip,\,Andre Zambanini}

\affiliation[a]{Peter Gr\"unberg Institute -- Integrated Computing Architectures (ICA | PGI-4), Forschungszentrum J\"ulich GmbH, Germany}
\affiliation[b]{Center for Particle Physics Siegen, Department Physik, Universit\"at Siegen, Germany}
\affiliation[c]{Elektronikentwicklungslabor, Department Physik, Universit\"at Siegen, Germany}
\emailAdd{dorosti@hep.physik.uni-siegen.de}
\affiliation[d]{Faculty of Engineering, Communication Systems,  University of Duisburg-Essen, Germany}

\abstract{\input{sections/00_abstract}}


\begin{document}
\maketitle

\flushbottom

\input{sections/01_introduction}
\FloatBarrier
\input{sections/02_data_sets_and_signal_construction}
\FloatBarrier
\input{sections/03_hybrid_trigger_method}
\FloatBarrier
\input{sections/04_trigger_performance}
\FloatBarrier
\input{sections/05_fpga_implementation}
\FloatBarrier
\input{sections/06_discussion}
\FloatBarrier
\input{sections/07_summary_and_outlook}
\FloatBarrier

\acknowledgments
\input{sections/08_acknowledgments}

\bibliographystyle{JHEP}
\bibliography{sections/09_literature}

\end{document}

%% file: sections/00_abstract.tex
Autonomous radio self-triggering for extensive air showers requires strong rejection of radio-frequency interference while preserving weak pulses within station-level hardware constraints. We present a hybrid neural trigger comprising a compact waveform denoiser followed by a classifier. The method is evaluated using experimentally measured high-interference background traces and detector-folded simulated air-shower pulses produced with the Pierre Auger \emph{Offline} chain, with the benchmark concentrated in the near-threshold regime. Hardware constraints are incorporated through hyperparameter optimisation, quantisation-aware training, and high-granularity fixed-point quantisation. Applying the conventional peak-envelope trigger after denoising increases its area under the receiver-operating-characteristic curve from \(0.63\) to \(0.98\). At a false-positive rate of \(10^{-4}\), the full denoiser--classifier chain retains about \(41\%\) of the held-out signal traces, compared with \(27\%\) for the classifier acting on the raw waveform and about \(2\%\) for the peak-envelope reference. The fixed-point firmware is synthesised, placed, and routed on representative field-programmable gate array targets, where it achieves timing closure with microsecond-scale latency and compact arithmetic-resource usage. These results establish hybrid neural denoising as a practical FPGA-compatible route toward radio-only triggering of weak and inclined air-shower signals in noisy environments.

%% file: sections/01_introduction.tex
\section{Introduction}
\label{sec:intro}

Ultra-high-energy cosmic rays (UHECRs) are studied with large ground-based observatories that combine particle detectors, fluorescence telescopes, and radio antennas~\cite{auger,TA_surface_detector,SCHRODER20171}. Radio detection has become particularly attractive because it offers a cost-effective technique with high duty cycle, broad sky coverage, and sensitivity to the electromagnetic component of extensive air showers (EAS)~\cite{HUEGE20161}. By exploiting the geomagnetic and charge-excess emission mechanisms~\cite{geomEffect,AskEffect}, radio arrays can reconstruct shower properties such as the primary energy and the depth of shower maximum with precision comparable to that achieved by established air-shower detection techniques~\cite{Abdul_Halim2024-ll,Aab2016-fq}.

One of the strongest motivations for radio-only triggering is the detection of inclined air showers. In that regime, the particle component is strongly attenuated before reaching ground level, whereas the radio pulse can still remain measurable over large footprints~\cite{Aab2016-fq}. Existing radio arrays therefore benefit from radio sensitivity precisely in the part of phase space where external particle-detector triggers become least efficient. A reliable self-trigger is consequently a key ingredient for extending the science reach of future cosmic-ray and neutrino observatories.

The long-standing obstacle is the radio background itself. In realistic field conditions, transient anthropogenic interference and narrow-band radio-frequency interference (RFI) generate large numbers of signal-like excursions that simple threshold triggers cannot reject efficiently~\cite{Schmidt2011-xy,Torres_Machado2013-my,KELLEY2013133}. A practical radio self-trigger must therefore do more than separate signal from background in principle: it must improve near-threshold selection while fitting within the latency and resource constraints of station-class field programmable gate array (FPGA) hardware.

A recent study by our group demonstrated a proof-of-principle classifier-only trigger trained on measured high-interference noise and simulated air-shower pulses, and showed that it outperforms a classical threshold trigger while preserving its performance after fixed-point conversion and FPGA-oriented synthesis~\cite{Dorosti:2025ugq}. That result established the feasibility of neural-network-based radio self-triggering in a realistic RFI environment, but it also left two natural next questions. First, can the trigger sensitivity be improved further, especially for signals near or below the operating threshold of the conventional peak-envelope trigger? Second, can the model-development strategy be changed such that the final implementation becomes substantially cheaper on the FPGA targets that matter for autonomous station-level deployment?

The present work addresses both questions by moving from a classifier-only trigger to a hybrid denoiser-plus-classifier pipeline and by making hardware constraints part of the model-development loop from the outset. The denoiser is introduced to recover weak pulses before the trigger decision, while hyperparameter optimisation (HPO), quantisation-aware training (QAT), and high-granularity quantisation (HGQ)~\cite{Sun2026-ce} are used to search for models that retain trigger performance at drastically reduced implementation cost. At the same time, the signal construction is made more realistic by replacing the earlier proof-of-principle simulation workflow with a detector-aware chain based on the Pierre Auger \emph{Offline} software framework~\cite{Argiro2007-gv}, while reusing the same measured-noise dataset as in the published study.

The Pierre Auger \emph{Offline} framework and Auger Engineering Radio Array (AERA) hardware provide the detector-response and station-scale instrumentation references used in this study~\cite{Maller2014AERA}. The proposed trigger is not specific to AERA and is evaluated here as a single-channel, fixed-window local selection block for autonomous radio stations.

The remainder of this paper is organised around these three developments. We first define the measured-noise benchmark, the detector-aware signal simulation, and the near-threshold signal construction used for the trigger study. We then introduce the denoiser--classifier architecture, the hyperparameter-optimisation procedure, and the QAT/HGQ workflow used to obtain firmware-compatible models. The trigger-performance analysis compares the denoiser-assisted branches with a classical peak-envelope threshold baseline under common fixed-false-positive operating conditions. The hardware study then evaluates the post-implementation resource and power demand of the resulting firmware and compares it with the previously published classifier-only FPGA implementation. Together, these studies quantify both the near-threshold trigger gain of the hybrid pipeline and the reduction in FPGA deployment cost achieved by the hardware-aware model-development workflow.

%% file: sections/02_data_sets_and_signal_construction.tex
\section{Data sets and signal construction}
\label{sec:data}

This section defines the measured background data, the detector-aware signal simulation, and the construction of the datasets used for the trigger study. The measured-noise traces and basic trace-conditioning convention follow the same experimental setup used previously, while the signal side is rebuilt around a detector-aware air-shower simulation based on the Pierre Auger \emph{Offline} framework~\cite{Argiro2007-gv}.

\subsection{Measured noise data}
\label{sec:noise}

We use the measured background dataset described in Ref.~\cite{Dorosti:2025ugq}. The noise traces were recorded on the physics campus of the University of Siegen using the same \SI{2}{\metre}$\times$\SI{2}{\metre} active Butterfly antenna and associated low-noise-amplifier hardware as deployed in the AERA Phase II stations~\cite{Charrier2012Butterfly,Weidenhaupt2015AERA}. The amplified signal was digitised at a sampling rate of \SI{250}{\mega\hertz} with an 8-bit digital oscilloscope. The digitised traces were processed offline with a fourth-order digital Butterworth band-pass filter to define the \SIrange{30}{80}{\mega\hertz} analysis band. The analogue out-of-band rejection of this acquisition chain was not characterised. The final model input consists of \num{128} samples, corresponding to an analysis window of \SI{512}{\nano\second} at this sampling rate.

The measured background corresponds to a difficult high-interference environment with both transient disturbances and prominent narrow-band RFI components. Keeping this background pool unchanged preserves a common and deliberately challenging reference point for all trigger branches. The acquisition setup, measurement campaign, and measured background spectrum, including the narrow-band and transient RFI components, are documented in Sec.~2 and Fig.~1 of Ref.~\cite{Dorosti:2025ugq}.

\subsection{Detector-aware signal simulation with the Pierre Auger \emph{Offline} framework}
\label{sec:simulation}

The signal model is based on a set of CoREAS/CORSIKA air-shower simulations~\cite{Heck1998CORSIKA} produced within the Pierre Auger Collaboration for energy-scale studies with the Auger Engineering Radio Array (AERA)~\cite{Fuchs2012AERA}; the showers were generated event by event following Ref.~\cite{Huege2025-dm}. The parent production is based on \num{844} high-quality AERA air showers with reconstructed surface-detector energies above \SI{3e17}{\electronvolt}. For each measured AERA event, two independent CORSIKA showers are generated using the measured geometry and calorimetric-energy estimate: one with a proton primary and one with an iron primary. The production described in Ref.~\cite{Huege2025-dm} uses Sibyll~2.3d and UrQMD as interaction models together with per-event atmospheric profiles derived from the Global Data Assimilation System.

The Pierre Auger \emph{Offline} chain provides a validated detector simulation including the antenna response and the subsequent electronic signal chain. In particular, the detector model used here includes the response of the AERA butterfly antennas together with the associated low-noise-amplifier electronics, so that the simulated detector output is matched to an experimentally calibrated radio-detector configuration. For the present work, we use the detector-folded voltage traces from this simulation chain as the starting point for signal injection. To match the acquisition system used for the measured-noise dataset of Sec.~\ref{sec:noise}, we retain only the low-gain signal chain, discard the additional amplification stage used in the AERA high-gain branch, and adapt the simulated traces to the \SI{250}{\mega\hertz} sampling used throughout this study.

The resulting library spans a broad range of shower geometries and pulse morphologies, inherited from the event-by-event AERA production and therefore covering variations in zenith angle, azimuth, $X_{\mathrm{max}}$, and distance from the antenna to the shower core. After the detector folding and adaptation to the local readout convention, we retain only traces in which the pulse reaches at least \num{50} analogue-to-digital converter (ADC) counts relative to the corresponding zero-crossing reference. This requirement suppresses effectively empty detector responses while keeping the signal library focused on traces that contain a resolved pulse candidate; lowering the retention value to \num{25} ADC counts was checked and did not materially change the benchmark conclusions. After the low-gain selection and the nominal \num{50}-ADC requirement, \num{9318} detector-folded traces remain in the retained signal library.

Combined with the measured background dataset of Sec.~\ref{sec:noise}, this construction yields signal-containing traces in which realistic detector-folded pulses are embedded into experimentally measured radio noise.

\subsection{Signal injection and amplitude scaling}
\label{sec:injection}

The detector-level pulses produced with the Pierre Auger \emph{Offline} chain are injected into measured noise traces at random temporal positions, using the same measured-noise pool described above. The construction is performed on traces of \num{160} samples, with the pulse maximum drawn uniformly between samples \num{20} and \num{134}. After filtering, \num{16} samples are removed from each edge to obtain the final \num{128}-sample model input, so that the pulse maximum ranges from samples \num{4} to \num{118} in the final trace. Because the filtered pulse has a finite duration and exhibits ringing, examples near these limits already include partial clipping by the final window boundaries; the edge margin therefore does not eliminate fixed-window boundary effects. This preserves the real transient and narrow-band interference structure of the recorded background while allowing controlled construction of signal-containing traces. Unless noted otherwise, noise-only and signal-containing traces are passed through the same digital conditioning chain, so that the trigger models never see differently processed background and signal samples.

In addition to the natural amplitude variation already present in the simulated event library, we apply a controlled post-simulation amplitude scaling in order to scan the near-threshold and sub-threshold regime with sufficient granularity. This scaling is not introduced as a replacement for the detector simulation, but as a practical way to probe trigger efficiency across a dense signal-strength range while keeping the detector-aware pulse shapes fixed. In that sense, the scaling step should be interpreted as part of the trigger-characterisation procedure rather than as an attempt to redefine the physical event distribution. The signal-containing samples used in the present study are intentionally concentrated more strongly in the weak-signal regime, because the central question is whether the denoiser--classifier pipeline improves performance close to and below the effective trigger threshold. Since this construction deliberately changes the signal-strength mixture, sample-integrated quantities such as the receiver-operating-characteristic (ROC) area under the curve (AUC) are benchmark-dependent. The main performance interpretation is therefore based on efficiencies at fixed false-positive rates (FPRs) and on their dependence on the signal-strength proxy, with an alternative global-scale signal construction used later as a robustness check of the amplitude-scaling procedure.

\begin{figure}[t]
\centering
\includegraphics[width=0.95\textwidth]{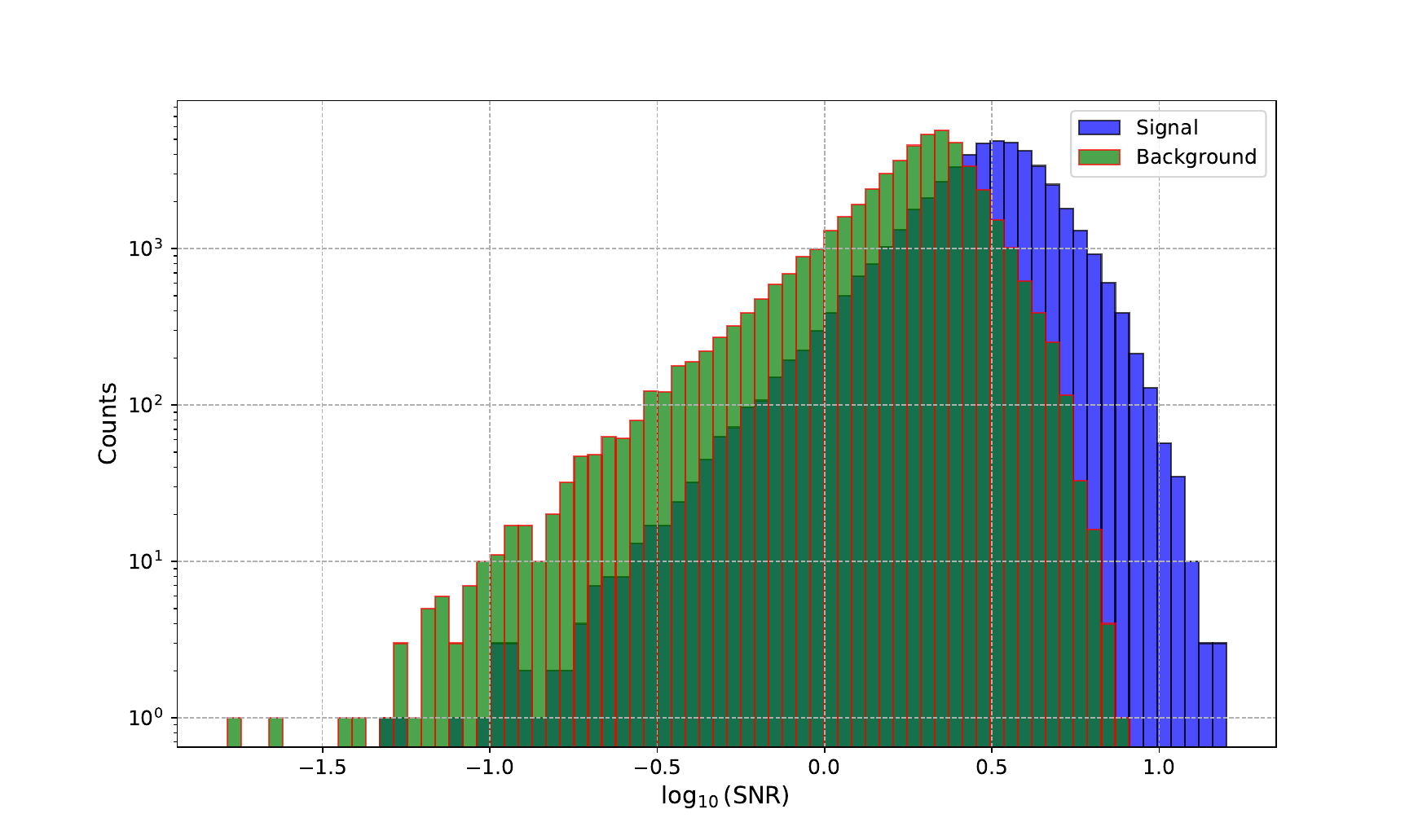}
\caption{Distribution of $\log_{10}(\mathrm{SNR})$ for the \texttt{signal} and \texttt{background} samples used in the present study, computed with the proxy defined in Eq.~\ref{eq:snr_proxy}. The figure is intended to show that the benchmark is concentrated strongly in the weak-signal regime relevant for the denoiser-assisted trigger.}
\label{fig:snr_distribution_shift}
\end{figure}

To characterise signal strength across the constructed datasets, we use a revised analysis-side signal-to-noise ratio (SNR) proxy for the histogram shown in figure~\ref{fig:snr_distribution_shift}. Let $\mathcal{P}$ denote the pulse window localised from the Hilbert envelope of the paired \texttt{pure signal} trace around its main peak, and let $|\mathcal{P}|$ denote the number of samples in this window. For a trace $x$, we then define
\begin{equation}
\label{eq:snr_proxy}
\mathrm{SNR}(x) = \frac{1}{\sigma_{\mathrm{noise}}^2}\left(\frac{1}{|\mathcal{P}|}\sum_{i\in\mathcal{P}} |\mathcal{H}(x)_i|^2\right),
\end{equation}
where $\mathcal{H}(x)$ denotes the Hilbert-envelope amplitude and $\sigma_{\mathrm{noise}}^2$ is the variance of the paired \texttt{background} trace. For the signal distribution, $x$ is the \texttt{signal} trace itself, so the plotted signal power is evaluated on the noisy waveform inside the pulse window. For the background distribution, the same pulse window is evaluated on the paired \texttt{background} trace. This differs from Ref.~\cite{Dorosti:2025ugq}, where the signal-side quantity was based on the clean-pulse peak amplitude and the background-side quantity was calculated directly from the noise trace itself. The revised proxy used here therefore tracks how strongly the pulse remains expressed inside the noisy waveform and provides a more direct signal-background comparison inside a common window. Qualitatively, the present benchmark should accordingly be read as more challenging at a given nominal SNR level than the benchmark of Ref.~\cite{Dorosti:2025ugq}, since the earlier definition was tied to the clean pulse rather than to its noisy realisation. The resulting $\log_{10}(\mathrm{SNR})$ distributions are shown in figure~\ref{fig:snr_distribution_shift}. They are intended to characterise the present benchmark and to show that it is populated much more strongly in the weak-signal regime, rather than to provide a one-to-one numerical comparison with the earlier histogram.

\subsection{Training, validation, and test samples}
\label{sec:samples}

Throughout the paper, the class \texttt{background} denotes measured noise-only traces, \texttt{pure signal} denotes detector-level simulated pulses without added noise, and \texttt{signal} denotes traces obtained by injecting a \texttt{pure signal} pulse into \texttt{background}. With this notation fixed, the denoiser is trained with \texttt{signal} traces as input and the corresponding \texttt{pure signal} traces as target, whereas the classifier is trained on \texttt{background} and \texttt{signal} traces. Hyperparameter optimisation is monitored on the \textbf{Cross-Validation Dataset}. For model development, the signal distribution is intentionally reweighted toward the near- and sub-threshold regime, while the noise distribution is kept unchanged.

The split is performed at the constructed-trace level. A constructed \texttt{signal} sample is defined by the detector-folded pulse, the measured background trace into which it is injected, the injection time, and the applied amplitude scale. Consequently, training and validation samples are disjoint as time-series traces, but they are drawn from the same parent library of detector-folded AERA simulations and from the same measured-noise campaign. The validation therefore tests generalisation to unseen pulse-noise realisations and amplitude scalings within this benchmark, rather than to a statistically independent shower-simulation production.

For the final physics and hardware comparisons, all methods are evaluated on held-out samples that are not used during model selection. \textbf{Validation Dataset 1} is the main signal-plus-background benchmark, and \textbf{Validation Dataset 2} is the large \texttt{background}-only sample used for false-positive studies. The same held-out samples are used throughout the ablation study of Sec.~\ref{sec:ablations}, so that the comparison remains one of trigger methods rather than of datasets.

Table~\ref{tab:dataset_splits} summarises the dataset definition used in the present study. The listed sizes correspond to the fixed split used throughout the analysis.

\begin{table}[t]
\centering
\small
\caption{Dataset definition used in the present study. The listed sample sizes and class compositions correspond to the fixed split used throughout the analysis.}
\label{tab:dataset_splits}
\begin{tabular}{>{\raggedright\arraybackslash}p{0.19\textwidth}>{\raggedright\arraybackslash}p{0.25\textwidth}r >{\raggedright\arraybackslash}p{0.27\textwidth}}
\toprule
Dataset & Current use in this work & Traces & Composition \\
\midrule
Training Dataset & Full model training & \num{150000} & \num{50000} \texttt{background}, \num{50000} \texttt{pure signal}, \num{50000} \texttt{signal} \\
Training Dataset & HPO training & \num{15000} & \textbf{denoiser:} \num{7500} \texttt{pure signal}, \num{7500} \texttt{signal}; \textbf{classifier:} \num{7500} \texttt{background}, \num{7500} \texttt{signal} \\
Training Dataset & HPO cross-validation & \num{10000} & \textbf{denoiser:} \num{5000} \texttt{pure signal}, \num{5000} \texttt{signal}; \textbf{classifier:} \num{5000} \texttt{background}, \num{5000} \texttt{signal} \\
Validation Dataset 1 & Full model validation & \num{150000} & \num{50000} \texttt{background}, \num{50000} \texttt{pure signal}, \num{50000} \texttt{signal} \\

Validation Dataset 2 & False-positive estimate & \num{1000000} & \texttt{background} only \\
\bottomrule
\end{tabular}
\end{table}

%% file: sections/03_hybrid_trigger_method.tex
\section{Hybrid trigger method}
\label{sec:architecture}

The present work moves from the classifier-only trigger of Ref.~\cite{Dorosti:2025ugq} to a deliberately co-designed pipeline in which waveform denoising, binary classification, and fixed-point deployment are treated as parts of a single trigger-development workflow. The underlying hypothesis is that weak air-shower pulses can be recovered more effectively if the trigger is allowed to perform a lightweight waveform-level cleaning step before the final signal-versus-background decision. At the same time, the model-development strategy is constrained from the outset by deployment requirements, so that improvements in trigger sensitivity are not obtained at the cost of impractical FPGA resource usage.

\subsection{Trigger concept and processing chain}
\label{sec:system_overview}

The trigger concept is a two-stage neural path operating on single-channel radio traces of length 128 samples. Let $\mathbf{x}\in\mathcal{X}\subset\mathbb{R}^{128}$ denote one input trace, where $\mathcal{X}$ is the bounded numerical domain defined by the preprocessing and numerical representation, and let $\theta$ and $\phi$ denote the trainable parameters of the denoiser and classifier, respectively. In the hybrid path, a denoiser $f_{\theta}$ first maps the noisy input waveform to a cleaned estimate,
\begin{equation}
\hat{\mathbf{x}} = f_{\theta}(\mathbf{x}),
\end{equation}
and a classifier $g_{\phi}$ then maps either the denoised trace $\hat{\mathbf{x}}$ or, in the classifier-only reference case, the raw input $\mathbf{x}$ to a signal probability
\begin{equation}
p_{\mathrm{sig}} = g_{\phi}(\hat{\mathbf{x}})\quad\text{or}\quad p_{\mathrm{sig}} = g_{\phi}(\mathbf{x}).
\end{equation}
A trigger is issued when $p_{\mathrm{sig}}$ exceeds an operating threshold $\tau$. This decomposition makes it possible to test directly whether a waveform-level denoising step adds value beyond what can already be achieved by a compact decision-level classifier.
Related convolutional-network studies have used this waveform-level perspective for radio-pulse identification and denoising in offline air-shower analyses~\cite{Schroder2024-ao,Abbasi2026-dn}; here, the same distinction between waveform recovery and decision making is used as a trigger-design question under fixed-point FPGA constraints.

The neural trigger path is compared primarily against a classical peak-envelope threshold trigger, which serves as the main non-neural baseline. This trigger operates on the peak amplitude of a signal-strength proxy and provides a conventional benchmark against which the gain of the denoiser-classifier pipeline can be evaluated. The previously published classifier-only model of Ref.~\cite{Dorosti:2025ugq} provides historical context for the development from a classifier-only trigger toward the denoiser-assisted hybrid pipeline.

From an implementation point of view, the full workflow proceeds from model development in \texttt{Keras} through quantisation-aware training and fixed-point export to firmware generation with \texttt{hls4ml}~\cite{fastml_hls4ml,Duarte:2018ite,Aarrestad:2021zos}, followed by high-level synthesis (HLS), register-transfer-level (RTL) generation, and integration of the resulting intellectual-property (IP) cores on the target FPGA. Figure~\ref{fig:workflow_diagram} separates the implemented inference path from this offline model-to-firmware and validation workflow.

\begin{figure}[t]
\centering
\begin{tikzpicture}[
    node distance=0.65cm,
    every node/.style={font=\scriptsize},
    block/.style={draw, thick, rounded corners, align=center,
                  minimum width=2.45cm, minimum height=0.85cm,
                  inner sep=3pt},
    flowblock/.style={draw, thick, rounded corners, align=center,
                      minimum width=2.35cm, minimum height=0.85cm,
                      inner sep=3pt},
    line/.style={-{Latex[length=2.2mm]}, thick},
    validation/.style={-{Latex[length=2.2mm]}, thick, dashed}
]

\node[font=\small\bfseries, anchor=west] (titlea) at (0,0)
      {(a) Implemented inference path};
\node[block, anchor=west] (input) at (0,-1.50)
      {Formed 128-sample window\\16-bit input stream};
\node[block, right=0.95cm of input] (denoiser) {Denoiser\\RTL IP};
\node[block, right=0.95cm of denoiser] (classifier) {Classifier\\RTL IP};
\node[block, right=0.95cm of classifier] (score)
      {Classifier score\\$p_{\mathrm{sig}}$};

\node[draw, thick, dashed, rounded corners,
      fit=(denoiser)(classifier), inner xsep=0.28cm, inner ysep=0.55cm,
      label={[font=\scriptsize, fill=white, inner sep=2pt]above:Target FPGA: \texttt{Vivado} block design}]
      (fpga) {};

\draw[line] (input) -- (denoiser);
\draw[line] (denoiser) -- (classifier);
\draw[line] (classifier) -- (score);

\node[font=\small\bfseries, anchor=west] (titleb) at (0,-3.70)
      {(b) Offline model-to-firmware and validation workflow};
\node[flowblock, anchor=west] (keras) at (0,-4.85)
      {Trained QAT/HGQ\\\texttt{Keras} models};
\node[flowblock, right=0.50cm of keras] (hls)
      {\texttt{hls4ml}\\fixed-point HLS/RTL};
\node[flowblock, right=0.50cm of hls] (vivado)
      {\texttt{Vivado}\\block design};
\node[flowblock, right=0.50cm of vivado] (implement)
      {Synthesis, placement,\\routing, timing closure};
\node[flowblock, right=0.50cm of implement] (reports)
      {Post-implementation\\resource/power results};

\node[flowblock, below=0.85cm of hls] (rtlval)
      {\texttt{cocotb} RTL simulation\\functional validation};

\draw[line] (keras) -- (hls);
\draw[line] (hls) -- (vivado);
\draw[line] (vivado) -- (implement);
\draw[line] (implement) -- (reports);
\draw[validation] (hls.south) -- (rtlval.north);

\end{tikzpicture}
\caption{Overview of the implemented FPGA trigger and the offline model-to-firmware workflow. (a) In the implemented inference path, one formed 128-sample input window is streamed through separate denoiser and classifier RTL IP cores connected within a single \texttt{Vivado} block design on the target FPGA, producing the classifier score $p_{\mathrm{sig}}$. (b) The trained QAT/HGQ \texttt{Keras} models are converted to fixed-point HLS/RTL IP with \texttt{hls4ml}, integrated in \texttt{Vivado}, and subjected to synthesis, placement and routing, timing closure, and post-implementation resource and power analysis. RTL simulation is used separately for functional validation against the fixed-point reference.}
\label{fig:workflow_diagram}
\end{figure}

\subsection{Denoiser and classifier model families}
\label{sec:denoiser}

The architecture is not presented as a single hand-designed network. Instead, the method begins from two compact model families, one for denoising and one for classification, which are intentionally restricted to low-complexity 1D convolutional forms suitable for later FPGA deployment.

The classifier family consists of shallow 1D convolutional networks that alternate convolution, non-linearity, and pooling before a low-dimensional decision head. The selected classifier is a compact six-block \texttt{Conv1D} network with \texttt{ReLU} activations and \texttt{MaxPool} stages, followed by global average pooling and a final dense output node. This model performs binary discrimination between \texttt{signal} and \texttt{background} traces and currently contains 460 trainable parameters.

The denoiser family is formulated as a lightweight fully convolutional regression model that maps a noisy trace to a cleaned waveform estimate. The selected denoiser uses four \texttt{Conv1D}+\texttt{ReLU} blocks together with an additive residual path, two intermediate \texttt{Conv1D}+\texttt{ReLU} layers, a further convolutional refinement stage, and a final single-filter output projection followed by reshaping back to a one-dimensional trace. This model currently contains 868 trainable parameters.

Using the notation fixed in Sec.~\ref{sec:samples}, the denoiser is trained as a waveform-regression model with \texttt{signal} traces as input and the corresponding \texttt{pure signal} traces as target, using a mean-squared-error loss. The classifier is trained with binary cross-entropy to distinguish \texttt{signal} from \texttt{background}. This notation is retained throughout the paper, so that the architectural description, the ablation study, and the hardware results all refer to the same sample definitions. Table \ref{tab:training_params} shows the key training parameters used for both models.

\begin{table}[t]
\centering
\caption{Training configuration for the denoiser and classifier models.}
\label{tab:training_params}
\begin{tabular}{lcc}
\hline
\textbf{Training Component} & \textbf{Denoiser} & \textbf{Classifier} \\
\hline
Loss function &
Mean squared error &
Binary cross-entropy \\[4pt]

Optimiser &
Adam &
Adam \\[4pt]

Learning-rate &
0.0016 &
0.0011 \\[6pt]

Batch size &
8 &
64 \\[4pt]

Stopping criterion &
Early stopping (patience = 50) &
Early stopping (patience = 20) \\[6pt]

\hline
\end{tabular}
\end{table}
The input traces are normalised for training using z-score normalisation. For each raw sample value $x_{\mathrm{raw}}$, the normalised value $x_{\mathrm{norm}}$ is defined as
\begin{equation}
    x_{\mathrm{norm}} = \frac{x_{\mathrm{raw}}-\mu}{\sigma},
\end{equation}
where $\mu$ and $\sigma$ denote the mean and standard deviation of the corresponding raw trace.

\begin{figure}[!htbp]
    \centering
    \includegraphics[width=1\linewidth]{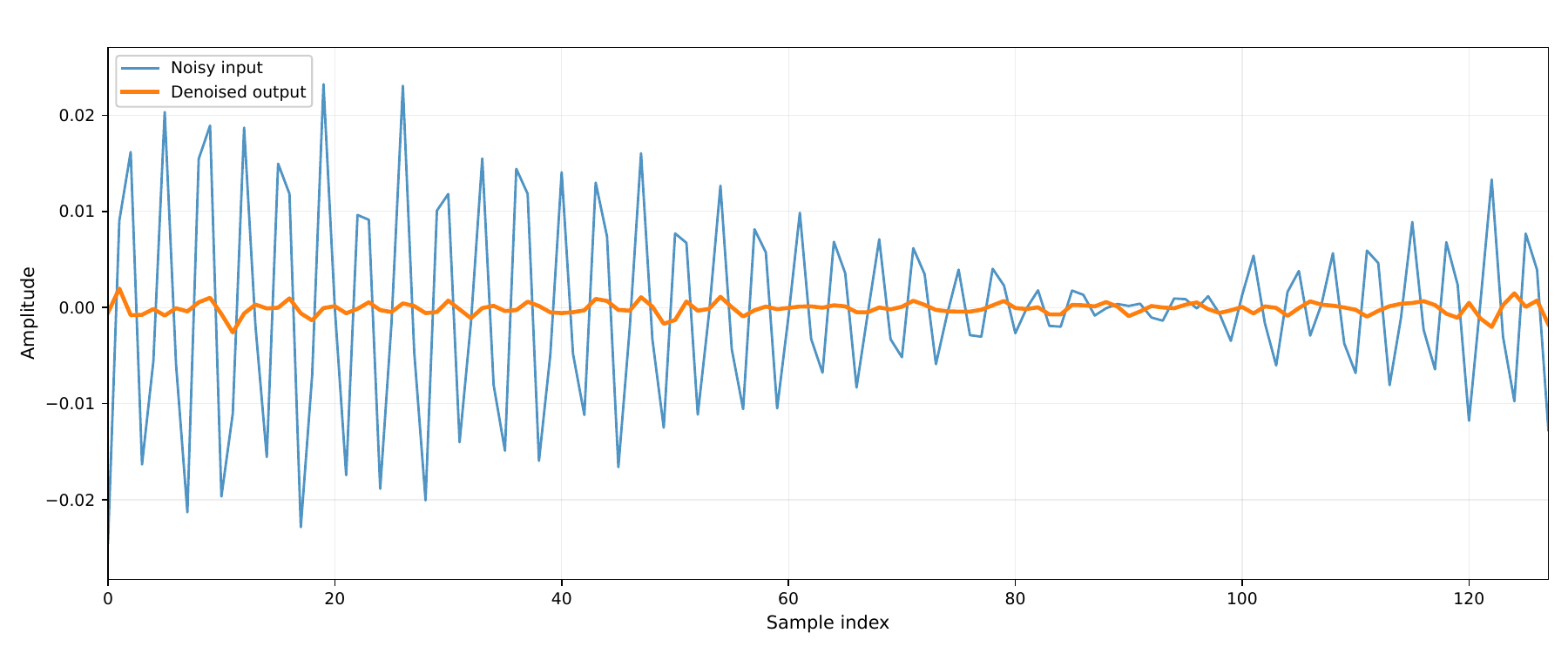}
    \caption{A denoised \texttt{background} (pure noise) trace. The blue curve shows the raw input trace, and the orange curve shows the denoised output. Since a \texttt{background} trace contains no simulated air-shower pulse, the target output is zero amplitude; the example illustrates that the denoiser suppresses the noise-only input without producing a signal-like pulse.}
    \label{fig:pure_noise}
\end{figure}

\begin{figure}[!htbp]
    \centering
    \includegraphics[width=1\linewidth]{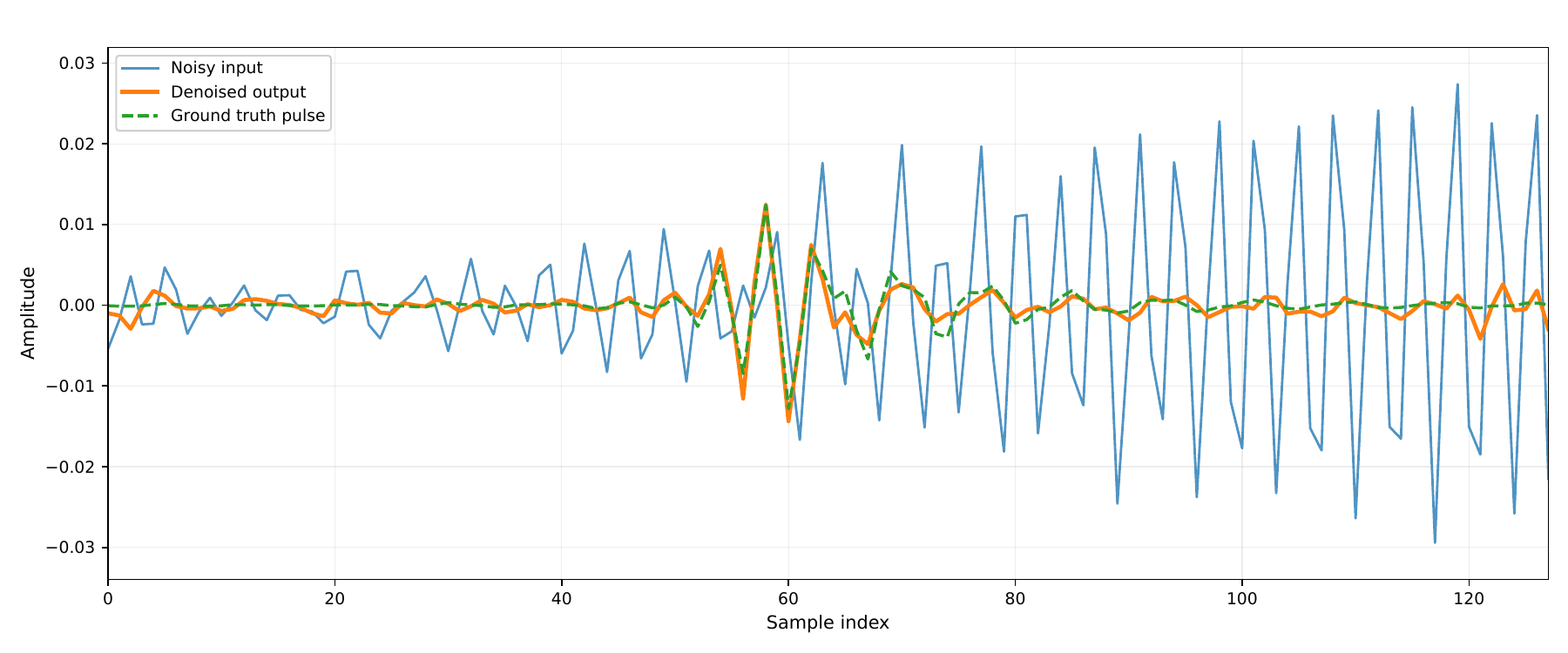}
    \caption{An example of a denoised \texttt{signal} trace. It can be seen that the signal is properly denoised with successful suppression of noise.}
    \label{fig:mixed_signal}
\end{figure}

Figures~\ref{fig:pure_noise} and~\ref{fig:mixed_signal} illustrate the denoiser response for representative \texttt{background} and \texttt{signal} traces. For the noise-only \texttt{background} case, where no physical pulse is present, the desired output is close to zero amplitude, whereas for the \texttt{signal} case the denoiser is expected to suppress noise while retaining the injected pulse structure.

\subsection{Hyperparameter optimisation setup and objective}
\label{sec:hpo}

Because even compact 1D convolutional models admit many architectural and training choices, model selection is automated with hyperparameter optimisation rather than fixed by manual trial and error. The current optimisation strategy is based on HyperBand~\cite{Li2018-hyperband}, which combines aggressive early stopping with successive halving. The basic logic is to train many candidate configurations for a small number of epochs, rank them after partial training, discard the least promising ones, and reallocate the available training budget to the surviving candidates. In this way, a comparatively large search space can be explored without fully training every candidate model.

For the trigger problem considered here, the natural search variables include the numbers of convolutional blocks, channel widths, kernel sizes, pooling pattern, residual connections or normalisation options, and training hyperparameters such as learning rate and batch size. The key methodological point is that model quality cannot be judged by classification accuracy alone. The search objective must also encode deployment realism. At the HPO stage, parameter count is used as an architecture-level proxy for hardware cost. No hard device-specific resource limit is applied to the classifier search. The soft penalty allows a larger model to be selected when its improvement in validation loss compensates for the additional parameter count. The actual resource and timing requirements are determined after FPGA implementation.

Table \ref{tab:hb} shows the HyperBand parameters used for the denoiser and classifier searches. The models have been optimised independently of each other.
\begin{table}[t]
\centering
\caption{HyperBand configuration used for model hyperparameter optimisation.}
\label{tab:hb}
\begin{tabular}{lcc}
\hline
\textbf{HyperBand Parameter} & \textbf{Denoiser} & \textbf{Classifier} \\
\hline
Maximum training budget &
$60$ epochs &
$50$ epochs \\[4pt]

Reduction factor &
$4$ &
$3$ \\[4pt]

Objective &
\multicolumn{2}{c}{minimise validation loss} \\[4pt]

Iterations &
$2$ &
$1$ \\[4pt]

Number of brackets &
$3$ &
$4$ \\[4pt]

Initial HPO candidates per bracket &
$3,6,16$ &
$4,6,12,27$ \\[4pt]

\hline
\end{tabular}
\end{table}
To make sure that HyperBand does not select models that are too costly for hardware deployment, the validation objective is constrained by model size. For the classifier, the validation binary cross-entropy (BCE) is augmented by a parameter-count penalty,
\begin{equation}
\mathcal{L}_{\mathrm{clf}} =
\mathrm{BCE}(y,\hat{y}) + 0.55\,\frac{N_{\mathrm{par}}}{2000},
\end{equation}
where $N_{\mathrm{par}}$ is the number of trainable parameters. The denominator \num{2000} sets a reference scale for compact classifier candidates and makes the parameter term dimensionless. With this choice, a \num{2000}-parameter model receives a loss penalty of \num{0.55}, while the selected \num{460}-parameter classifier receives a penalty of \num{0.13}. The coefficient was chosen as an empirical compactness prior for the HyperBand ranking, so that larger models are accepted only if their validation loss improves enough to compensate for the additional parameters.

For the denoiser, the search uses a hard model-size constraint rather than a soft penalty. Candidate architectures above the sub-\num{1000}-parameter target are assigned a constant loss of $10^9$ and are therefore discarded by the HyperBand selection. Valid denoiser candidates are ranked by their validation mean-squared-error loss. HyperBand has been used on HGQ models~\cite{Sun2026-ce} in the case of the classifier and QKeras~\cite{Coelho2021-rr} for the denoiser (see section~\ref{sec:qat_hgq}).

The result of this optimisation and selection procedure is a compact classifier and a compact denoiser, which together define the model set used in the remainder of the paper. The classifier represents the decision stage against which the denoiser-assisted trigger can be compared, while the denoiser represents the waveform-cleaning stage selected for the deployment-oriented studies. Figure \ref{fig:architectures} shows the selected architectures together with their key hyperparameters.


\tikzset{
  den_arrow/.style={
    postaction={
      decorate,
      decoration={
        markings,
        mark=at position 0.05 with {\arrow{latex}}
      }
    }
  }
}

\tikzset{
  class_arrow/.style={
    postaction={
      decorate,
      decoration={
        markings,
        mark=at position 0.07 with {\arrow{latex}}
      }
    }
  }
}

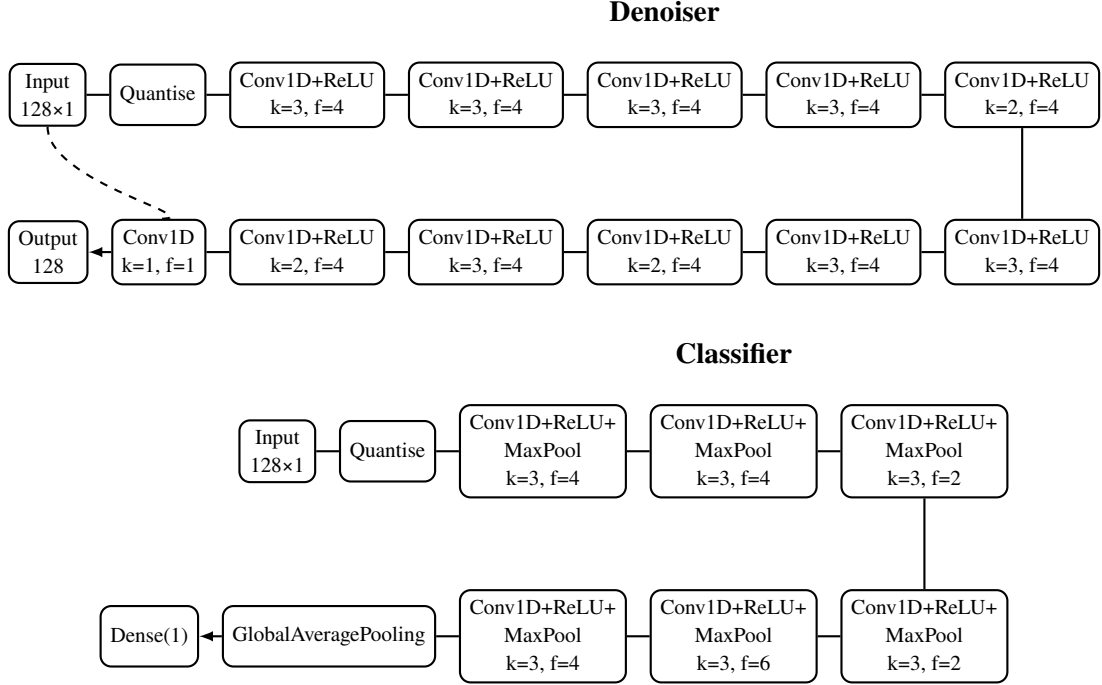
\begin{figure}[!htbp]
\centering

\begin{tikzpicture}[
    box/.style={draw, thick, rounded corners, align=center,
                minimum width=1.0cm, minimum height=0.8cm,
                font=\scriptsize},
    line/.style={-latex, thick},
    skip/.style={dashed, thick},
    node distance=0.3cm
]

\node[box] (in) {Input\\128×1};
\node[box, right=0.3cm of in] (q) {Quantise};

\node[box, right=0.3cm of q] (e1) {Conv1D+ReLU\\k=3, f=4};
\node[box, right=0.3cm of e1] (e2) {Conv1D+ReLU\\k=3, f=4};
\node[box, right=0.3cm of e2] (e3) {Conv1D+ReLU\\k=3, f=4};
\node[box, right=0.3cm of e3] (e4) {Conv1D+ReLU\\k=3, f=4};

\node[box, right=0.3cm of e4] (b1) {Conv1D+ReLU\\k=2, f=4};
\node[box, below=1.2cm of b1] (b2) {Conv1D+ReLU\\k=3, f=4};

\node[box, left=0.3cm of b2] (d4) {Conv1D+ReLU\\k=3, f=4};
\node[box, left=0.3cm of d4] (d3) {Conv1D+ReLU\\k=2, f=4};
\node[box, left=0.3cm of d3] (d2) {Conv1D+ReLU\\k=3, f=4};
\node[box, left=0.3cm of d2] (d1) {Conv1D+ReLU\\k=2, f=4};

\node[box, left=0.3cm of d1] (res) {Conv1D\\k=1, f=1};
\node[box, left=0.3cm of res] (out) {Output\\128};

\draw[line] (in) -- (q) -- (e1) -- (e2) -- (e3) -- (e4)
            -- (b1) -- (b2) -- (d4) -- (d3) -- (d2) -- (d1)
            -- (res) -- (out);

\draw[skip] (in.south) to[out=-90,in=0] (res.north);

\node[font=\bfseries, above=0.4cm of e3] {Denoiser};

\end{tikzpicture}

\vspace{0.6cm}

\begin{tikzpicture}[
    box/.style={draw, thick, rounded corners, align=center,
                minimum width=1.0cm, minimum height=0.8cm,
                font=\scriptsize},
    line/.style={-latex, thick},
    node distance=0.3cm
]

\node[box] (c_in) {Input\\128×1};
\node[box, right=0.3cm of c_in] (c_q) {Quantise};

\node[box, right=0.3cm of c_q] (c1) {Conv1D+ReLU+\\MaxPool\\k=3, f=4};
\node[box, right=0.3cm of c1] (c2) {Conv1D+ReLU+\\MaxPool\\k=3, f=4};
\node[box, right=0.3cm of c2] (c3) {Conv1D+ReLU+\\MaxPool\\k=3, f=2};
\node[box, below=1.2cm of c3] (c4) {Conv1D+ReLU+\\MaxPool\\k=3, f=2};
\node[box, left=0.3cm of c4] (c5) {Conv1D+ReLU+\\MaxPool\\k=3, f=6};
\node[box, left=0.3cm of c5] (c6) {Conv1D+ReLU+\\MaxPool\\k=3, f=4};

\node[box, left=0.3cm of c6] (gap) {GlobalAveragePooling};
\node[box, left=0.3cm of gap] (dense) {Dense(1)};

\draw[line] (c_in) -- (c_q) -- (c1) -- (c2) -- (c3)
            -- (c4) -- (c5) -- (c6) -- (gap) -- (dense);

\node[font=\bfseries, above=0.4cm of c2] {Classifier};

\end{tikzpicture}

\caption{Architecture diagrams of the denoiser (top) and classifier (bottom). The kernel size is abbreviated by $k$, the number of filters by $f$. The dashed line denotes a skip connection.}
\label{fig:architectures}
\end{figure}


\subsection{Quantisation-aware training and high-granularity quantisation}
\label{sec:qat_hgq}

Deployment compatibility is addressed already during training rather than only after the floating-point models are fixed. Quantisation-aware training (QAT) is the central mechanism for this step: the models are trained under fixed-point constraints so that the optimisation sees, at least approximately, the numerical limitations that will later be imposed by the FPGA implementation. This reduces the mismatch between software training and hardware inference and provides a more reliable basis for converting the selected models with \texttt{hls4ml} into synthesizable firmware~\cite{fastml_hls4ml,Duarte:2018ite,Aarrestad:2021zos}.\\
\noindent \\
\textbf{Classifier:} The classifier is trained using HGQ, which adapts the precisions of weights and activations during quantisation-aware training. For each HGQ layer $l$, a differentiable estimate of its bit-operation count, $\mathrm{BOPs}_l$, is included in the training objective,
\[
\mathcal{L}_{\mathrm{HGQ}}
=
\mathcal{L}_{\mathrm{task}}
+
\beta\sum_l \mathrm{BOPs}_l .
\]
The fixed hyperparameter $\beta$ therefore controls the strength of the hardware-complexity penalty: increasing $\beta$ favours lower learned precisions, while decreasing it gives greater weight to the task loss. We use $\beta=8\times10^{-6}$ for all HGQ layers of the classifier. The resulting fixed-point precisions used for HLS are listed in Table \ref{tab:prec_classifier}. The parameter $\beta$ is used only during training and does not enter the FPGA inference calculation.\\
\noindent \\
\textbf{Denoiser:} Here, the bit widths have been set globally to \texttt{ap\_fixed<14,8,RND,SAT>}, with a $\beta=8\cdot10^{-7}$, as summarised in Table \ref{tab:prec_denoiser}.\\
\noindent \\
Both the classifier and denoiser are quantised independently and have been trained quantisation-aware. The hardware impact of the quantisation strategy is quantified in Sec.~\ref{sec:fpga}, where the post-implementation resource use, latency, and power of the complete trigger firmware are reported.

\begin{table}[h!]
\centering
\small
\caption{Final fixed-point precisions used for each classifier block in the high-level synthesis (HLS) implementation. ReLU denotes rectified linear unit, and GAP denotes global average pooling.}
\label{tab:prec_classifier}
\begin{tabular}{l l l c c}
\hline
\textbf{Block} & \textbf{Type} & \textbf{Layer} & \textbf{Total bits} & \textbf{Integer bits} \\
\hline
1 & Weight & Conv1D & 5 & 2 \\
1 & Bias   & Conv1D & 4 & 1 \\
1 & Result & Conv1D & 14 & 4 \\
1 & Result & ReLU   & 11 & 4 \\
1 & Result & MaxPool   & 11 & 4 \\
\hline
2 & Weight & Conv1D & 6 & 2 \\
2 & Bias   & Conv1D & 7 & 1 \\
2 & Result & Conv1D & 17 & 6 \\
2 & Result & ReLU   & 13 & 5 \\
2 & Result & MaxPool   & 13 & 5 \\
\hline
3 & Weight & Conv1D & 6 & 2 \\
3 & Bias   & Conv1D & 4 & 1 \\
3 & Result & Conv1D & 16 & 6 \\
3 & Result & ReLU   & 14 & 6 \\
3 & Result & MaxPool   & 14 & 6 \\
\hline
4 & Weight & Conv1D & 5 & 2 \\
4 & Bias   & Conv1D & 7 & 1 \\
4 & Result & Conv1D & 16 & 7 \\
4 & Result & ReLU   & 14 & 7 \\
4 & Result & MaxPool   & 14 & 7 \\
\hline
5 & Weight & Conv1D & 7 & 2 \\
5 & Bias   & Conv1D & 7 & 2 \\
5 & Result & Conv1D & 19 & 8 \\
5 & Result & ReLU   & 15 & 7 \\
5 & Result & MaxPool   & 15 & 7 \\
\hline
6 & Weight & Conv1D & 6 & 2 \\
6 & Bias   & Conv1D & 7 & 1 \\
6 & Result & Conv1D & 19 & 8 \\
6 & Result & ReLU   & 15 & 7 \\
6 & Result & MaxPool   & 15 & 7 \\
\hline
Head & Result & GAP    & 15 & 7 \\
Head & Weight & Dense  & 6 & 1 \\
Head & Bias   & Dense  & 6 & 1 \\
Head & Result & Dense  & 15 & 6 \\
\hline
\end{tabular}
\end{table}

\begin{table}[h!]
\centering
\caption{Global fixed-point precision used for the denoiser HLS implementation.}
\label{tab:prec_denoiser}
\begin{tabular}{l c c}
\hline
\textbf{Denoiser precision policy} & \textbf{Total bits} & \textbf{Integer bits} \\
\hline
Global \texttt{ap\_fixed<14,8,RND,SAT>} precision & 14 & 8 \\
\hline
\end{tabular}
\end{table}

%% file: sections/04_trigger_performance.tex
\section{Trigger performance}
\label{sec:performance}

This section compares the trigger branches considered in the present study against the classical threshold reference. The internal logic follows the model variants introduced by the current development workflow: a denoiser-only branch, a classifier-only branch, and the full denoiser--classifier pipeline. Because the signal-containing samples are intentionally shifted toward the near- and sub-threshold regime, all comparisons must be made on the held-out benchmark samples defined in Sec.~\ref{sec:samples}.

\subsection{Evaluation metrics and operating points}
\label{sec:metrics}

The primary performance metric is the receiver-operating-characteristic (ROC) curve, expressed in terms of the true-positive rate (TPR) for \texttt{signal} traces and the false-positive rate (FPR) for \texttt{background} traces. The area under the curve (AUC) is reported as a compact summary, but the practically relevant comparison is made at fixed operating points, since a station trigger is never operated over the full ROC. Throughout this section, \textbf{Validation Dataset 1} provides the main signal-plus-background benchmark, while \textbf{Validation Dataset 2} provides the large \texttt{background}-only sample needed for robust low-FPR estimates.

The same operating points are used for all trigger branches: the threshold reference, the denoiser-only branch, the classifier-only branch, and the full denoiser--classifier chain.
Unless stated otherwise, the neural ROC curves in this section refer to the quantisation-aware trained HGQ models evaluated in the \texttt{Keras} environment. The corresponding firmware-exported \texttt{hls4ml} response is shown explicitly for the full chain as a conversion consistency check.
Here, a low FPR$_{\mathrm{low}}$ of $10^{-4}$ has been chosen as a benchmark for very strict false-positive requirements, corresponding to $0.01 \%$ of evaluated noise-only windows misidentified as signals. The nominal benchmark operating point is FPR$_{\mathrm{nom}} = 10^{-2}=1 \%$. These FPR values are defined per evaluated input window. For a candidate-window rate $R_{\mathrm{cand}}$, the corresponding noise-trigger rate is $R_{\mathrm{cand}}\times\mathrm{FPR}$. As an operating reference, AERA stations transmitted approximately \num{500} local T2 trigger timestamps per second to the central data-acquisition system, which reduced them through array-level selection to approximately \SIrange{2}{5}{\hertz} T3 requests~\cite{Maller2014AERA}.

\subsection{Classical threshold baseline}
\label{sec:threshold}

The classical reference is the same peak-envelope threshold trigger already used in Ref.~\cite{Dorosti:2025ugq}. For each trace, the analytic signal is formed and the trigger score is derived from the Hilbert-envelope peak together with the corresponding noise normalisation, following the published definition. The threshold branch is evaluated on the same band-pass-filtered traces as the neural branches, so the comparison is performed on one common input representation.

This threshold trigger serves two purposes in the present paper. First, it provides the conventional reference against which the neural methods are compared. Second, it provides the decision stage for the denoiser-only branch described below, where the same peak-envelope threshold is applied after waveform cleaning.

\subsection{Comparison of trigger branches}
\label{sec:ablations}
The following trigger branches are compared to each other:
\begin{enumerate}
\item \textbf{Denoiser-only}: the trace is first denoised, and the decision is then made by the peak-envelope threshold of Sec.~\ref{sec:threshold};
\item \textbf{Classifier-only}: the classifier acts directly on the band-pass-filtered raw trace;
\item \textbf{Denoiser + classifier (hybrid trigger)}: the denoised trace is passed to the classifier.
\end{enumerate}

These three branches isolate the two possible sources of improvement: the denoiser-only branch tests whether waveform cleaning helps a conventional threshold decision, while the classifier-only and denoiser--classifier branches test whether the denoised waveform adds information beyond a direct decision-level classifier. Table~\ref{tab:ablation_comparison} summarises the final branch-level trigger comparison. It compares the branch logic itself, separating waveform cleaning from decision-level classification while keeping all operating points common across the threshold baseline, denoiser-only branch, classifier-only branch, and full denoiser--classifier chain.

\begin{table}[t]
\centering
\small
\caption{Main branch-level trigger comparison. The neural branches improve the trigger classification compared to the baseline peak-envelope trigger by increasing the ROC AUC and the TPR at the nominal and low-FPR operating points.}
\label{tab:ablation_comparison}
\begin{tabular}{>{\raggedright\arraybackslash}p{0.28\textwidth}ccc}
\toprule
Trigger branch & AUC & TPR@FPR$_{\mathrm{nom}}$ & TPR@FPR$_{\mathrm{low}}$ \\
\midrule
Threshold baseline & 0.629 & 0.1 & 0.02 \\
Denoiser-only & 0.98 & 0.87 & 0.16 \\
Classifier-only & 0.984 & 0.84 & 0.27 \\
Denoiser + classifier & 0.986 & 0.89 & 0.41 \\
\bottomrule
\end{tabular}
\end{table}

\begin{figure}[!htbp]
    \centering
    \includegraphics[width=1\linewidth]{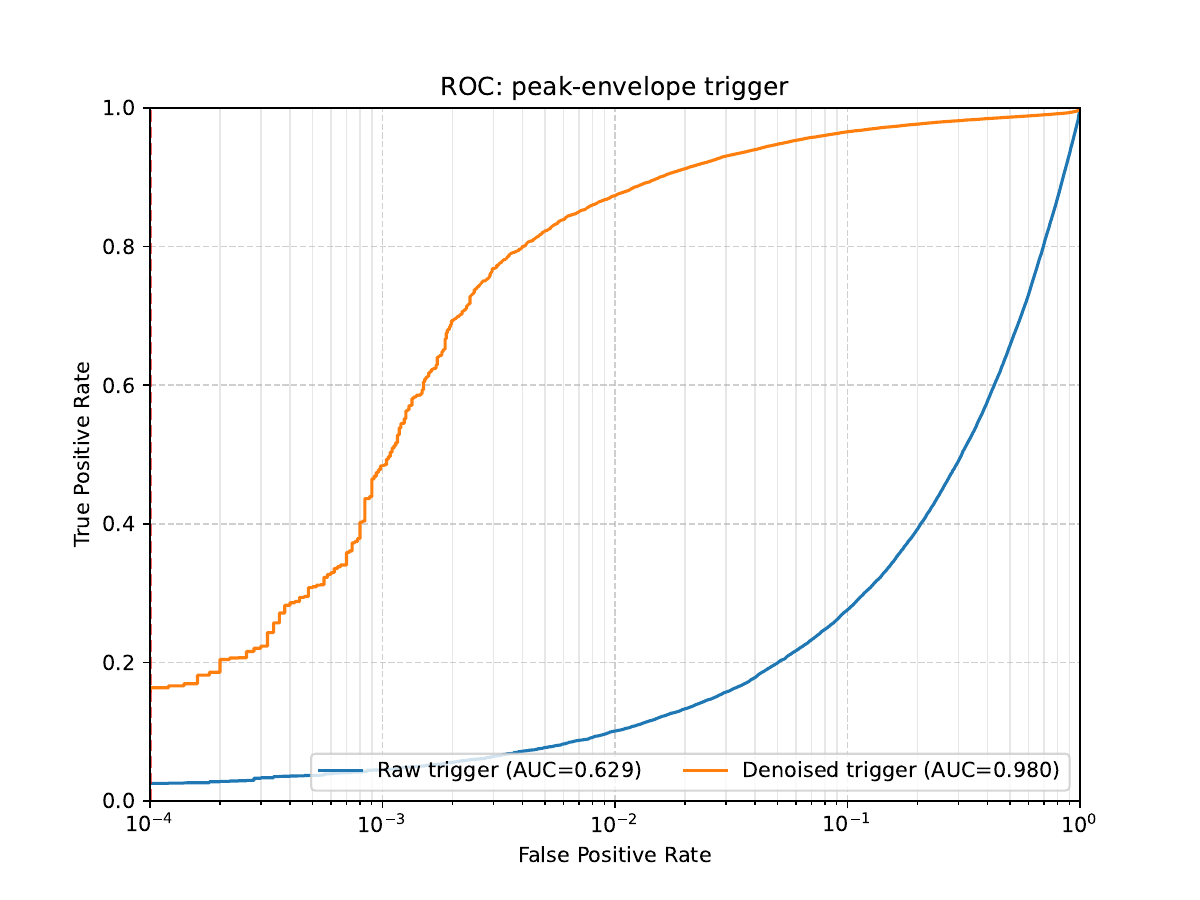}
    \caption{ROC curve of the denoiser-only branch compared with the peak-envelope threshold applied to the raw trace. Applying the same threshold after denoising increases the AUC from 0.63 to 0.98. The evaluation uses all signal and background traces from Validation Dataset~1, whose signal-strength distribution is shown in Fig.~\ref{fig:snr_distribution_shift}.}
    \label{fig:roc_denoiser}
\end{figure}

The ROC figures are organised in the same order as the branch logic. Figure~\ref{fig:roc_denoiser} shows whether denoising alone improves the separation obtained with a simple peak-envelope threshold. A peak-envelope trigger that acts on the raw noisy input achieves an AUC of 0.63; after denoising, this value increases to 0.98.

\begin{figure}[!htbp]
    \centering
    \includegraphics[width=1\linewidth]{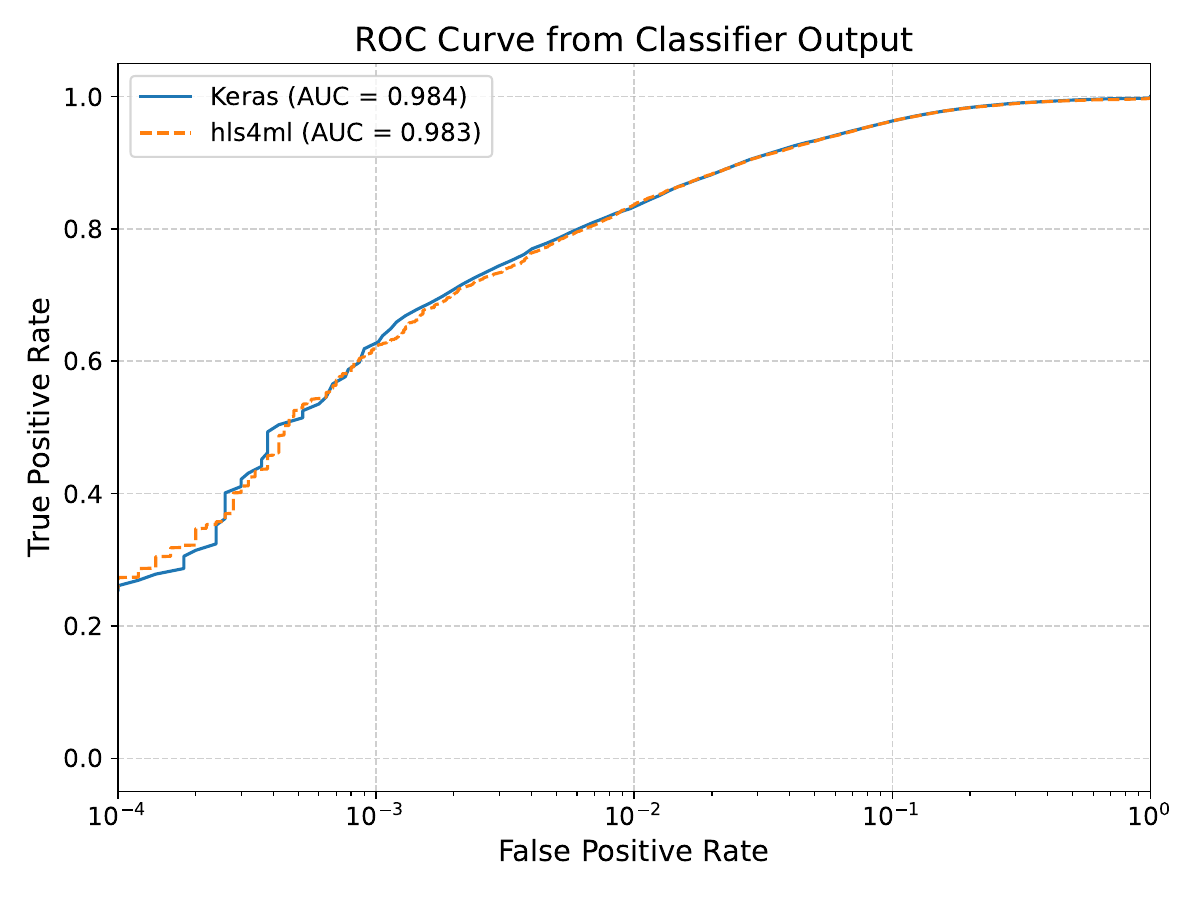}
    \caption{ROC curve of the classifier-only branch. The HGQ/\texttt{Keras} classifier reaches an AUC of 0.984 and maintains signal efficiency at lower false-positive rates than the classical threshold trigger.}
    \label{fig:roc_classifier}
\end{figure}

Figure~\ref{fig:roc_classifier} shows the ROC curve of the classifier-only branch. The AUC increases to 0.984 compared to the classical peak-envelope trigger. The classifier also retains signal efficiency in the low-FPR region, where the classical trigger retains only about \(2\%\) of signal traces at FPR \(=10^{-4}\) in the present benchmark.

\begin{figure}[!htbp]
    \centering
    \includegraphics[width=1\linewidth]{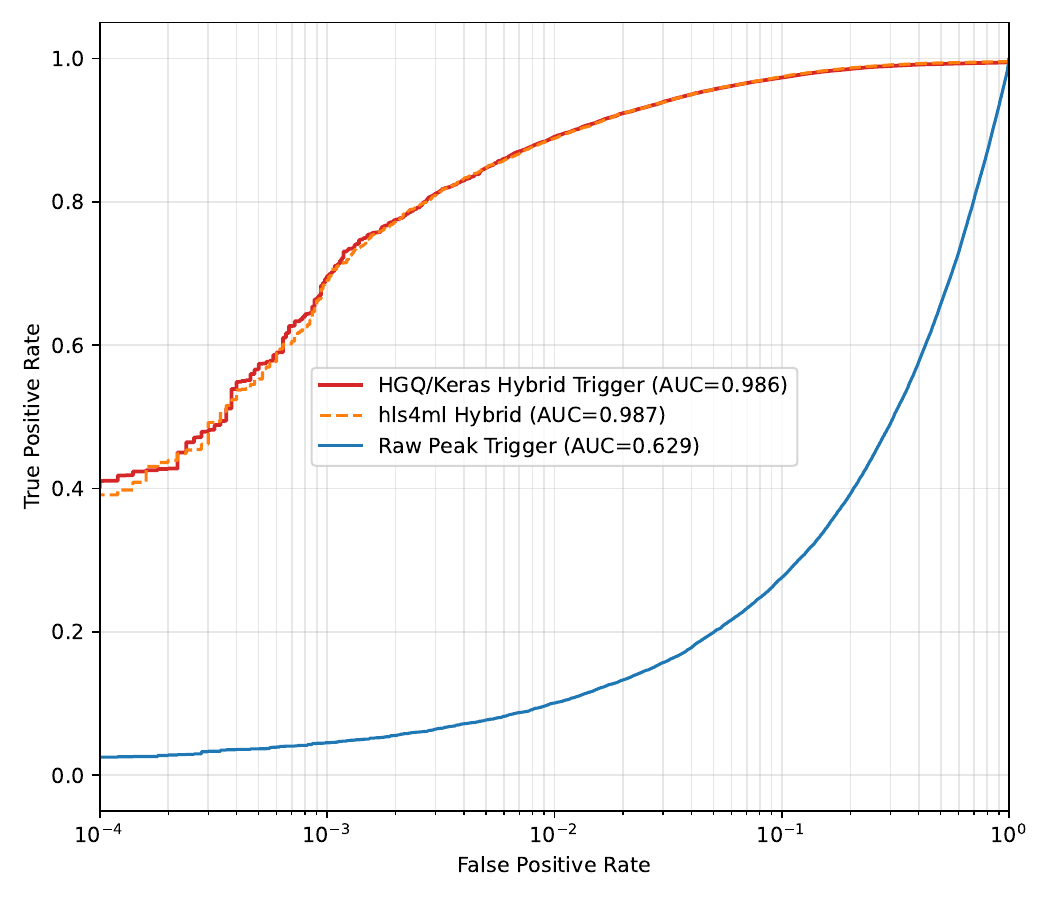}
    \caption{ROC curve of the full denoiser--classifier chain (hybrid trigger) compared with the peak-envelope threshold applied to the raw trace, shown in blue. The HGQ/\texttt{Keras} branch reaches an AUC of 0.986 and gives a TPR of about 0.41 at FPR$_{\mathrm{low}}=10^{-4}$. The \texttt{hls4ml} export closely follows the HGQ/\texttt{Keras} curve, with an AUC of 0.987 and a slightly lower TPR of about 0.39 at FPR$_{\mathrm{low}}=10^{-4}$.}
    \label{fig:roc_both}
\end{figure}

Figure~\ref{fig:roc_both} shows the ROC curve of the full denoiser--classifier chain, which we refer to as the hybrid trigger in the following. The hybrid branch reaches an AUC of 0.986 and retains a TPR of about 0.41 at the strict operating point FPR$_{\mathrm{low}}=10^{-4}$. The low-FPR part of the curve shows a plateau near this efficiency before rising toward higher false-positive rates. The \texttt{hls4ml} curve closely follows the HGQ/\texttt{Keras} result, with a slightly larger AUC of 0.987 and a TPR of about 0.39 at FPR$_{\mathrm{low}}=10^{-4}$; this small difference is interpreted as a conversion-level effect rather than a real performance gain. Together, these figures show the progression from classical triggering to waveform cleaning and finally to the full hybrid trigger.

\begin{figure}[!htbp]
    \centering
    \includegraphics[width=1\linewidth]{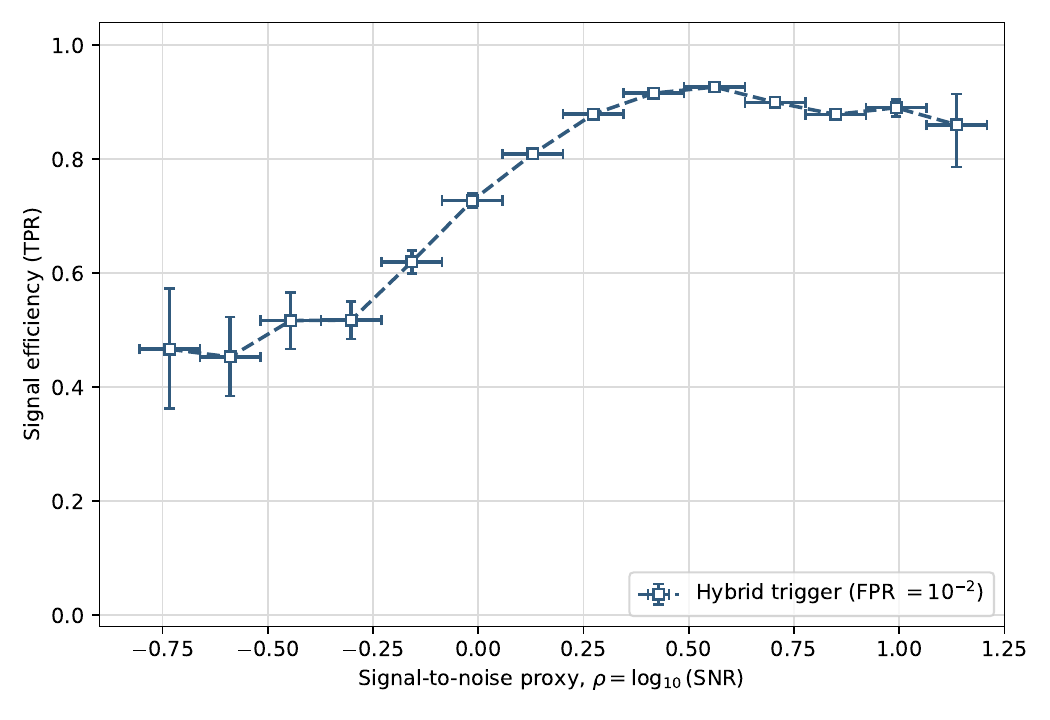}
    \caption{Signal efficiency as a function of the SNR proxy for the hybrid trigger at the nominal operating point FPR$_{\mathrm{nom}}=10^{-2}$, evaluated on the signal traces of Validation Dataset~1. The signal-to-noise conditions are characterised using the detector-level proxy $\rho=\log_{10}(\mathrm{SNR})$ defined in Eq.~\ref{eq:snr_proxy}. Vertical bars show 68\% Clopper--Pearson intervals, and horizontal bars indicate the bin widths.}
    \label{fig:hybrid_turn_on}
\end{figure}

To resolve the hybrid-trigger response across signal-to-noise regimes, figure~\ref{fig:hybrid_turn_on} shows the signal efficiency as a function of $\rho=\log_{10}(\mathrm{SNR})$ at the nominal benchmark operating point, FPR$_{\mathrm{nom}}=10^{-2}$. Using the SNR proxy defined in Eq.~\ref{eq:snr_proxy}, $\rho$ characterizes the signal-to-noise conditions of each waveform, accounting for contributions from both the injected pulse and the measured background noise. The efficiency shown in figure~\ref{fig:hybrid_turn_on} is therefore conditional on this proxy. The small decrease at high $\rho$ reflects differences among the pulse--noise pairs populating those bins, rather than a decrease in average efficiency as pulse amplitude increases. When the same validation pulses are rescaled while their noise traces are held fixed, the efficiency increases at every tested amplitude and approaches saturation.

The ablation therefore leads to two conclusions. Denoising alone already makes a simple peak-envelope trigger substantially more efficient, demonstrating that the waveform-cleaning stage preserves trigger-relevant signal structure. The full denoiser--classifier chain gives the largest overall AUC and the strongest performance at both the nominal and strict low-FPR operating points, and is therefore the preferred branch for the background-limited trigger regime. The low-FPR operating point is evaluated with the large \texttt{background}-only sample of \textbf{Validation Dataset 2}, so that the strict false-positive comparison is not limited by the smaller mixed validation sample.

%% file: sections/05_fpga_implementation.tex
\section{FPGA implementation}
\label{sec:fpga}

This section presents the FPGA realisation of the hybrid trigger and quantifies the resulting hardware cost. The key outcome is not only that the model can be synthesised, but that the new pipeline changes the deployment frontier compared with the previously published design. The FPGA implementation uses the QAT/HGQ models described in Sec.~\ref{sec:qat_hgq}, whose trigger-level agreement with the fixed-point \texttt{hls4ml} representation was verified in figure~\ref{fig:roc_both}; this section therefore focuses on resource use, timing, latency, and power.

\subsection{Target platforms and tool flow}
\label{sec:platforms}

The hardware study follows the same benchmarking logic as the classifier-only implementation in Ref.~\cite{Dorosti:2025ugq}. The same fixed-point RTL design is synthesised, placed, and routed on three representative FPGA targets. The Zynq-7000 Z-7020 provides a constrained station-class feasibility test, while the larger targets show how resource mapping, timing closure, and latency change across device classes. This comparison concerns the implementation of one fixed design on different devices; model-size scaling is not studied here. The Z-7020 should therefore be read as a stress test of the implementation, not as a prescription for the final deployment device. The resulting resource and timing summary is given in Table~\ref{tab:fpga_results}.

The trained QAT/HGQ denoiser and classifier are converted from the \texttt{Keras} model description to a fixed-point HLS representation with \texttt{hls4ml}~\cite{fastml_hls4ml,Duarte:2018ite,Aarrestad:2021zos}. The generated HLS design is synthesised with \texttt{Vitis HLS~2024.1}~\cite{vitisHLS2023}, targeting the representative FPGA platforms listed in Table~\ref{tab:fpga_results}. For the post-implementation study, the FPGA timing constraint was relaxed to \SI{6}{\nano\second} to obtain robust timing closure on the most constrained target, the Zynq-7000 Z-7020, using the unmodified register-transfer-level (RTL) design generated by \texttt{hls4ml}. This implementation clock only sets the internal processing rate used for hardware timing and latency estimates; it does not change the physics sampling assumptions of the input traces. Since the remaining timing pressure is marginal, further timing-oriented HLS or RTL pipelining would be expected to reduce this constraint, but such device-specific optimisation is outside the scope of the present comparison. The following subsections report the hardware cost of the complete denoiser--classifier trigger chain; any differences between the HGQ/\texttt{Keras} and \texttt{hls4ml} responses have already been checked at trigger level in Sec.~\ref{sec:ablations}.

After HLS synthesis, the denoiser and classifier are exported as separate RTL intellectual-property (IP) blocks and assembled in a \texttt{Vivado~2024.1} block design~\cite{vivado2024}. The design contains the two \texttt{hls4ml}-generated IP cores, the top-level wiring between them, and the clock, reset, start, and ready/control logic required to exercise the chain. The denoiser IP receives the 16-bit input stream and its output is connected to the classifier input, so that the implemented design corresponds to the complete denoiser--classifier trigger path. The resource, latency, timing, and power values reported below are obtained after Vivado synthesis and implementation, including placement, routing, and timing closure for each target device. For the standalone hardware test design, the input stream is generated by a 16-bit pseudo-random binary sequence (PRBS) source. The PRBS block is implemented as a linear-feedback shift register initialised to all ones. On each clock cycle, the register is shifted by one bit and the feedback bit is
\[
b_{\mathrm{fb}} = b_{15}\oplus b_{13}\oplus b_{12}\oplus b_{10},
\]
corresponding to taps of the polynomial $x^{16}+x^{14}+x^{13}+x^{11}$. This source is not part of the physics trigger chain itself and is not used for the end-to-end RTL validation in Sec.~\ref{sec:rtl_validation}. It is included only in the standalone implementation setup used to drive non-constant activity when assessing post-implementation resource use and dynamic power. The PRBS excitation produces deliberately high switching activity. The reported dynamic-power values therefore provide conservative high-activity estimates, with lower values expected for typical traces and constant inputs.

\subsection{Resource use, timing, and power}
\label{sec:resources}

Table~\ref{tab:fpga_results} summarises the post-implementation results for the complete denoiser--classifier trigger chain on the representative FPGA targets. Placed-and-routed designs include routing, packing, and timing-closure effects that are not visible in high-level synthesis estimates alone, and are therefore the relevant quantities for assessing implementation feasibility.

\begin{table}[t]
\centering
\small
\caption{Post-implementation resource use, latency, and timing closure for the full denoiser--classifier design on representative FPGA targets. LUT denotes lookup-table resources, FF denotes flip-flop resources, BRAM denotes block random-access memory, and DSP denotes digital-signal-processing blocks. The Z-7020 is used as the conservative low-resource reference, while the larger devices show scaling behaviour and timing headroom.}
\label{tab:fpga_results}
\resizebox{\textwidth}{!}{%
\begin{tabular}{lcccccc}
\toprule
Platform & LUT & FF & BRAM & DSP & Latency [$\mu$s] & Timing met \\
\midrule
Zynq-7000 Z-7020 & \num{22906} & \num{39817} & \num{50} & \num{42} & \num{13.55} & yes \\
Kintex UltraScale xcku040 & \num{22624} & \num{25370} & \num{31.5} & \num{68} & \num{5.66} & yes \\
Alveo U250 xcu250 & \num{22287} & \num{20777} & \num{32} & \num{70} & \num{3.63} & yes \\
\bottomrule
\end{tabular}%
}
\end{table}

The Z-7020 row is the most conservative one: its resource use should be interpreted as implementation pressure on a constrained reference device, not as a universal resource estimate for the model. Conversely, the larger targets provide additional DSP, memory, and routing headroom, which is reflected in the reduced latency and continued timing closure. The end-to-end latency decreases from \SI{13.55}{\micro\second} on the Z-7020 to \SI{5.66}{\micro\second} on the Kintex UltraScale target and \SI{3.63}{\micro\second} on the Alveo U250, while all three implementations meet the imposed timing constraint after placement and routing.

With the initiation interval equal to the reported latency, the slowest implementation, on the Z-7020, can accept approximately $7.4\times10^{4}$ windows per second. Direct evaluation of every non-overlapping \num{128}-sample window in a \SI{250}{\mega\hertz} stream would require approximately $1.95\times10^{6}$ evaluations per second and therefore at least 27 parallel Z-7020 instances. A single-instance implementation consequently requires an upstream candidate generator. At the nominal FPR of $10^{-2}$, limiting the station-level noise-trigger output to the AERA operating scale of approximately \SI{500}{\hertz} requires a candidate-window rate no larger than approximately \SI{50}{\kilo\hertz}, which is within the processing capacity of one instance. This rate relation defines a system-level requirement; the resulting FPR must ultimately be validated on the background distribution produced by the upstream candidate generator.

The quoted dynamic-power values used for the comparison with the previous FPGA study are discussed in Sec.~\ref{sec:fpga_baseline}. These values are obtained from the post-implementation FPGA power analysis using the activity model of the implemented design, providing a consistent design-level comparison across the target devices.

\subsection{Hardware-cost reduction relative to the published FPGA study}
\label{sec:fpga_baseline}

Table~\ref{tab:fpga_baseline_comparison} compares the present design with the previously published FPGA implementation of Ref.~\cite{Dorosti:2025ugq}. The resource comparison is made for the same target FPGA, the Zynq-7000 Z-7020. For the published design, however, the quoted resource numbers correspond to the post-synthesis report, because that design could not be placed and routed on the Z-7020. In contrast, the current design reaches implementation on the Z-7020. The dynamic-power estimates are therefore listed separately for the Z-7020 and for the Alveo U250, so that the station-class implementation result and the direct comparison to the published Alveo implementation are both visible.

\begin{table}[t]
\centering
\small
\caption{Comparison of FPGA resource demand and dynamic-power estimates for the published FPGA implementation and the current deployment-oriented hybrid trigger design. Resource numbers are quoted for the Zynq-7000 Z-7020 target in both cases. The published design could not be implemented on the Z-7020, so no Z-7020 dynamic-power estimate is available for that row; the U250 column gives the direct power comparison to the device used in Ref.~\cite{Dorosti:2025ugq}.}
\label{tab:fpga_baseline_comparison}
\resizebox{\textwidth}{!}{%
\begin{tabular}{llccccc}
\toprule
Design & Resource report & DSP & LUT & BRAM & Z-7020 dyn. power [W] & U250 dyn. power [W] \\
\midrule
Published FPGA study~\cite{Dorosti:2025ugq} & post-synthesis on Z-7020 & 2119 & 191243 & 780 & -- & 5.51 \\
Current design & post-implementation on Z-7020 & 42 & 22920 & 50 & 0.562 & 0.270 \\
\bottomrule
\end{tabular}%
}
\end{table}

The most important change is the DSP demand on the Z-7020 target, which drops from 2119 blocks at post-synthesis level in the published study to 42 blocks after implementation in the current design. Since DSP availability was the dominant bottleneck in the earlier FPGA study, this reduction directly changes the feasibility of deploying the trigger on constrained station-level hardware. The dynamic-power comparison on the Alveo U250, the device used for the published implementation, decreases from \SI{5.51}{\watt} to \SI{0.270}{\watt}. The current design also implements directly on the Z-7020, where the post-implementation dynamic-power estimate is \SI{0.562}{\watt}. LUT and BRAM usage follow the same downward trend, supporting the conclusion that the present design is not merely a performance extension of the previous study, but a substantially cheaper hardware implementation.

For context, the digital electronics developed for autonomous AERA radio stations, including digitisation and local FPGA trigger processing, were reported to consume approximately \SI{10.8}{\watt}~\cite{Maller2014AERA}. The \SI{0.562}{\watt} high-activity dynamic-power estimate of the present Z-7020 implementation is therefore approximately $5\%$ of this station-electronics scale, placing integration of the trigger logic within reach of a solar-powered autonomous radio station.

\subsection{End-to-end RTL validation}
\label{sec:rtl_validation}

Beyond synthesis and implementation reports, the design response is validated at the RTL level with a Python-based \texttt{cocotb} testbench~\cite{Gadde2024-rl}. The purpose of this step is to verify that the generated hardware logic preserves the model response when driven with representative traces, rather than only checking that the design can be synthesised. The testbench applies the same input normalisation used in the software validation, evaluates the quantised reference model at Python/\texttt{hls4ml} level, drives the generated hardware design in simulation, and records the resulting RTL trigger probabilities.

The RTL simulation is computationally more expensive than the software-level studies, so the end-to-end comparison is performed on a balanced validation subset containing \num{2000} signal traces and \num{2000} background traces. The absolute AUC values obtained on this subset should therefore not be compared one-to-one with the full-statistics curves in Sec.~\ref{sec:performance}. The relevant validation here is instead the comparison between the \texttt{hls4ml} reference and the RTL output on the same events.

\begin{figure}[t]
\centering
\includegraphics[width=0.95\textwidth]{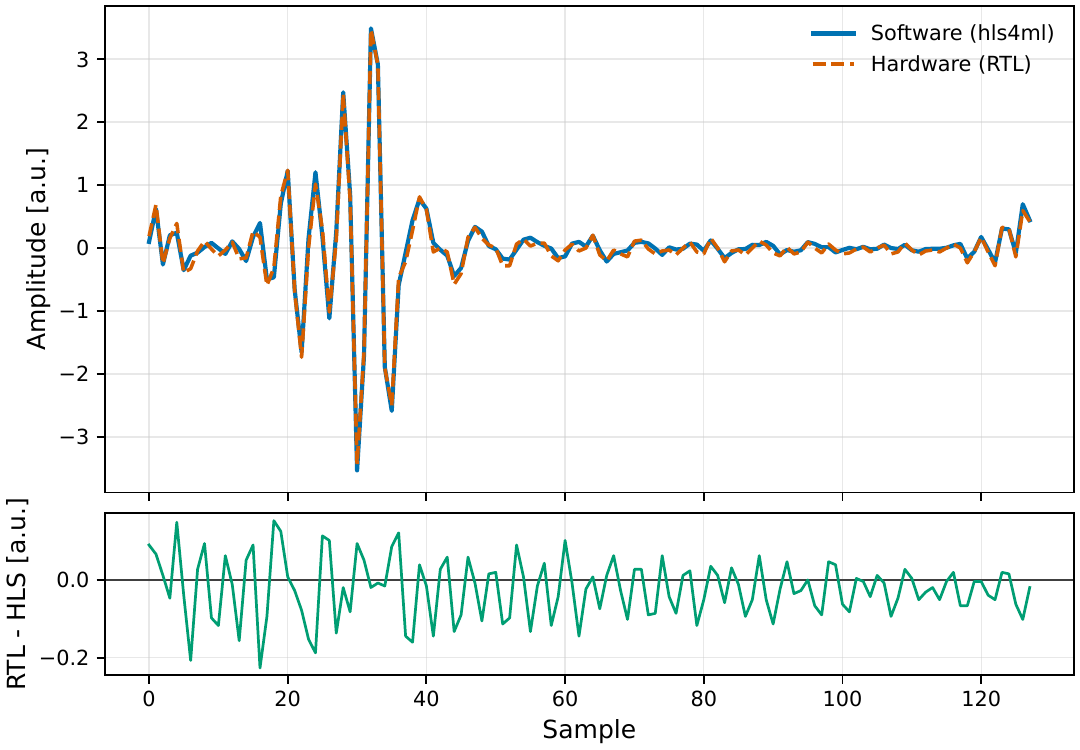}
\caption{Event-level RTL validation of the denoiser output for a representative trace. The \texttt{hls4ml} reference and RTL waveform agree with a correlation coefficient of 99.5\% for the example shown; the lower panel shows the residual between the RTL and \texttt{hls4ml} outputs.}
\label{fig:rtl_denoiser_trace}
\end{figure}
\FloatBarrier

Figure~\ref{fig:rtl_denoiser_trace} first checks the denoiser output at waveform level. For the representative trace shown, the \texttt{hls4ml} and RTL outputs have a correlation coefficient of 99.5\%, and the residual remains small compared with the coherent pulse amplitude. This event-level check verifies that the RTL simulation of the generated design preserves the denoised trace shape that is passed to the classifier. Across the complete RTL validation subset of \num{2000} signal-containing and \num{2000} background traces, the median per-trace RMS residual is 0.0977 and 0.0886\,a.u., respectively. For signal-containing traces, the residual RMS has only a weak correlation with the normalised pure-signal peak ($r=0.14$) and no observable correlation with the normalised noise RMS ($r=-0.007$). The RTL--\texttt{hls4ml} difference therefore remains approximately constant in absolute amplitude over the tested range and does not scale appreciably with the signal or noise amplitude.

\begin{figure}[t]
\centering
\includegraphics[width=0.78\textwidth]{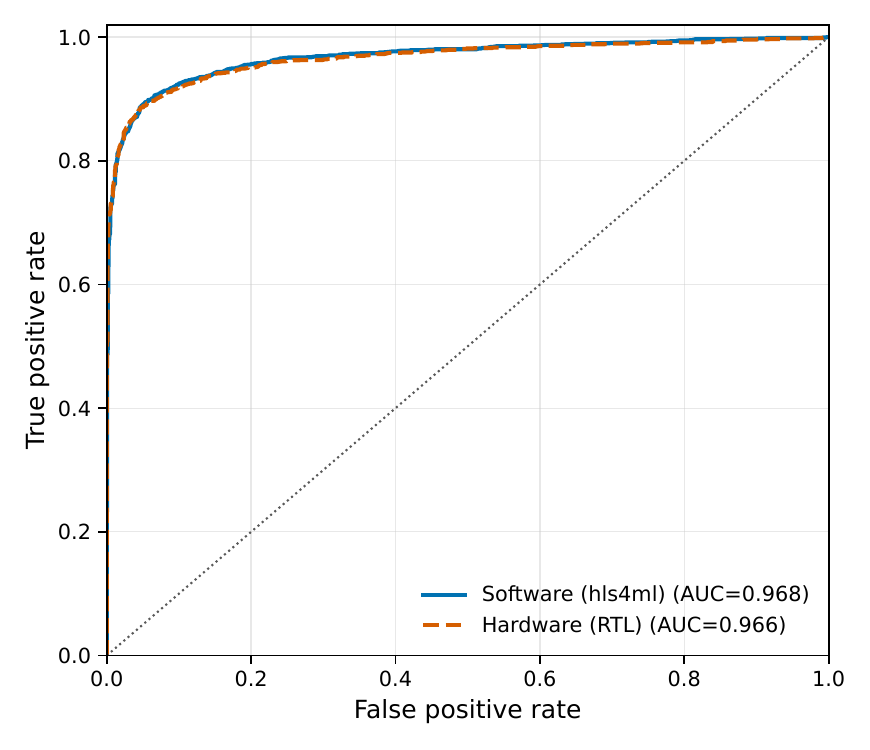}
\caption{ROC-level validation of the end-to-end RTL simulation against the quantised \texttt{hls4ml} reference on the balanced RTL validation subset containing \num{2000} signal traces and \num{2000} background traces. The software reference gives an AUC of 0.968, while the RTL simulation gives an AUC of 0.966.}
\label{fig:rtl_roc}
\end{figure}

The trigger-level validation is shown in figure~\ref{fig:rtl_roc}. The ROC curves from the RTL simulation and the \texttt{hls4ml} reference nearly overlap over the full false-positive-rate range, with AUC values of 0.966 and 0.968, respectively. This small difference is consistent with the finite validation subset and the numerical details of the RTL realisation. More importantly, the agreement is evaluated event-by-event on the same traces, showing that the hardware-realised trigger preserves the signal-background ordering of the quantised software reference. The floating-point and HGQ/\texttt{Keras} comparisons are not repeated here, since their agreement with the \texttt{hls4ml} model has already been established in Sec.~\ref{sec:ablations}.

%% file: sections/06_discussion.tex
\section{Discussion}
\label{sec:discussion}

The trigger and hardware results address the same instrumentation question from complementary directions: whether waveform denoising can improve autonomous radio triggering without exceeding the resource envelope of an edge FPGA implementation. The discussion therefore focuses on the gain in recoverable weak traces at fixed background load, the additional information carried by the denoised waveform beyond a decision-level score, and the deployment steps needed to translate the demonstrated station-level trigger block into a broader radio-array readout strategy.

\subsection{Physics and operational implications}
\label{sec:implications}

The trigger comparison developed in Sec.~\ref{sec:ablations} is most directly interpreted through the recovery of weak signal-containing traces at fixed background load. This is the operating regime that limits autonomous radio triggering, because conservative thresholds suppress precisely the low-amplitude pulses that are most vulnerable to noise fluctuations. For this purpose, we compare the hybrid trigger directly to the threshold baseline. The signal-strength coordinate is the same revised analysis-side proxy introduced in Eq.~\ref{eq:snr_proxy}. Writing
\begin{equation}
\rho \equiv \log_{10}(\mathrm{SNR}),
\end{equation}
we bin the held-out \textbf{Validation Dataset 1} in $\rho$ and evaluate, in identical bins, the signal efficiencies of the threshold branch and the hybrid branch. The operating thresholds of the two branches are chosen independently on held-out \texttt{background} traces so that both correspond to the same target false-positive rate. This yields two directly comparable efficiency curves, $\epsilon_{\mathrm{base}}(\rho)$ and $\epsilon_{\mathrm{hyb}}(\rho)$, under matched background conditions.

The most compact physics-facing summary is then the conditional detection gain
\begin{equation}
G(\rho) \equiv \frac{\epsilon_{\mathrm{hyb}}(\rho)}{\epsilon_{\mathrm{base}}(\rho)}.
\end{equation}
By construction, $G(\rho) > 1$ identifies signal-strength bins in which the hybrid trigger recovers traces that would be missed by the classical threshold trigger at the same background load. Bins in which the baseline efficiency is consistent with zero do not admit a stable finite ratio and should therefore not be overinterpreted.

\begin{figure}[t]
\centering
\IfFileExists{figures/gain_vs_snr.pdf}{%
  \includegraphics[width=0.95\textwidth]{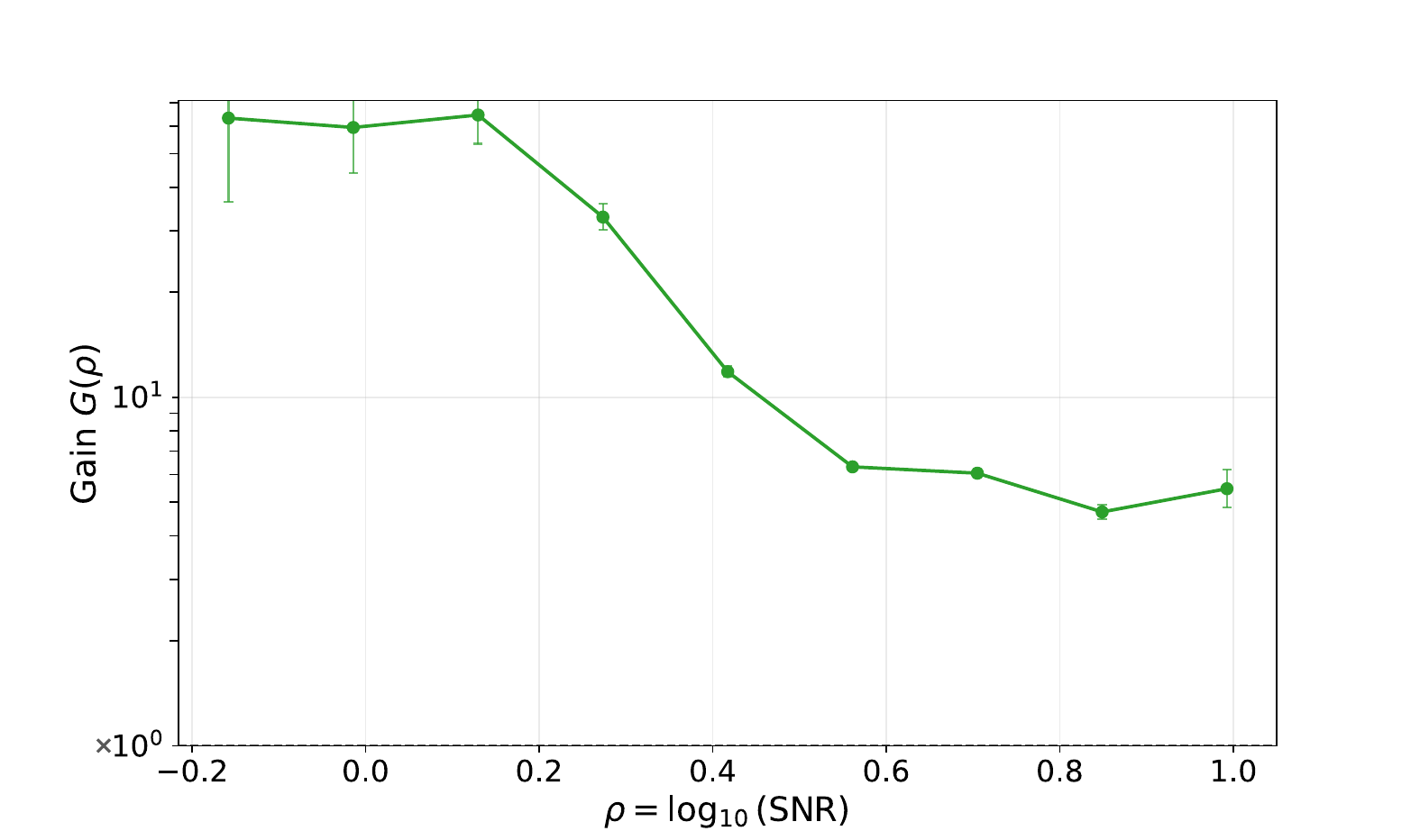}%
}{%
  \fbox{%
    \parbox[c][5.2cm][c]{0.72\textwidth}{%
      \centering
      Conditional gain $G(\rho)$ vs.\ $\rho$\\[0.5em]
      hybrid trigger relative to threshold baseline\\
      at fixed target FPR
    }%
  }%
}
\caption{Efficiency gain $G(\rho)=\epsilon_{\mathrm{hyb}}(\rho)/\epsilon_{\mathrm{base}}(\rho)$ as a function of the signal-strength proxy $\rho=\log_{10}(\mathrm{SNR})$ for the hybrid trigger relative to the threshold baseline, evaluated at a fixed false-positive rate of $10^{-2}$. The curve quantifies the relative increase in detectable signal-containing traces in the near-threshold regime.}
\label{fig:gain_vs_snr_discussion}
\end{figure}

Figure~\ref{fig:gain_vs_snr_discussion} therefore answers the station-level trigger question directly: within the validated signal-strength range of the benchmark, how much extra near-threshold radio content becomes available to a radio-only trigger path when the hybrid trigger is applied?

Within this formulation, the physics implication is immediate. Any systematic increase in $G(\rho)$ over a finite low-$\rho$ interval corresponds to a proportional increase in the number of detectable signal-containing traces in precisely the regime that limits autonomous radio triggering. The hybrid trigger therefore relaxes an instrumental limitation that otherwise forces conservative thresholds and suppresses weak events.

At the fixed false-positive operating point of $10^{-2}$, the conditional detection gain is substantial throughout the weak-signal regime. In the lowest finite bin shown in figure~\ref{fig:gain_vs_snr_discussion}, centred at $\rho=-0.158$, the gain is $G=63.2^{+52.1}_{-26.8}$, where the asymmetric uncertainty follows from Clopper--Pearson intervals propagated to the efficiency ratio. Even at the upper end of the plotted finite range, $\rho=0.993$, the gain remains $G=5.46^{+0.73}_{-0.64}$. These values show that the hybrid trigger provides a strong multiplicative increase in recoverable near-threshold events, rather than a marginal improvement, precisely in the regime that limits autonomous radio triggering.

This same low-$\rho$ gain is the part of the result with the clearest relevance for inclined and otherwise radio-faint geometries. Such events deposit less per-station amplitude and are therefore disproportionately exposed to efficiency losses from conservative thresholding. A trigger branch that preserves more efficiency in the weak-signal bins at fixed background rate removes part of that bias at the station level, even before any array-level acceptance study is attempted.

\subsection{What the denoiser adds beyond a decision-level trigger}
\label{sec:why_denoiser}

The classifier-only branch is a strong benchmark because it already uses waveform information before making a binary decision. The question is therefore not whether a neural decision stage is useful, but whether an explicit waveform-regression stage adds information that cannot be reduced to another decision-level score. The ablation in Table~\ref{tab:ablation_comparison} gives the most direct answer. At the nominal operating point, the classifier-only and full denoiser--classifier branches have comparable efficiencies, with TPR values of $0.84$ and $0.89$, respectively. The denoiser is therefore not introduced because the classifier-only branch fails at loose background conditions, but because it improves the trigger response across the benchmark while also providing a cleaned waveform. At FPR$_{\mathrm{low}}=10^{-4}$, the classifier-only branch reaches a TPR of $0.27$, while the full denoiser--classifier branch reaches about $0.41$.

This behaviour is consistent with the denoiser-only branch. If the same peak-envelope threshold is applied after waveform cleaning, the AUC increases from $0.63$ for the raw threshold trigger to $0.98$ for the denoised threshold trigger. This is an important control test because the downstream decision rule is then deliberately simple. The improvement cannot be attributed to a more flexible classifier head; it shows that the denoiser changes the waveform representation itself by suppressing noise excursions while preserving pulse-like structure.

The denoiser also provides a diagnostic object that a pure classifier does not: a cleaned trace that can be inspected, thresholded, compressed, or passed to other station-level logic. This is relevant beyond the single-station trigger score. In a bandwidth-limited radio array, the station does not only need to decide whether a trace is interesting; it may also need to transmit a compact set of features, such as pulse time, peak amplitude, or integrated envelope, to support array-level coincidence building, direction reconstruction, quality cuts, or selective waveform readout. A classifier-only trigger can provide a probability, but it does not by itself produce a waveform from which these features can be extracted with a simple and auditable edge algorithm.

\begin{figure}[t]
\centering
\begin{subfigure}{0.48\textwidth}
    \centering
    \includegraphics[width=\linewidth]{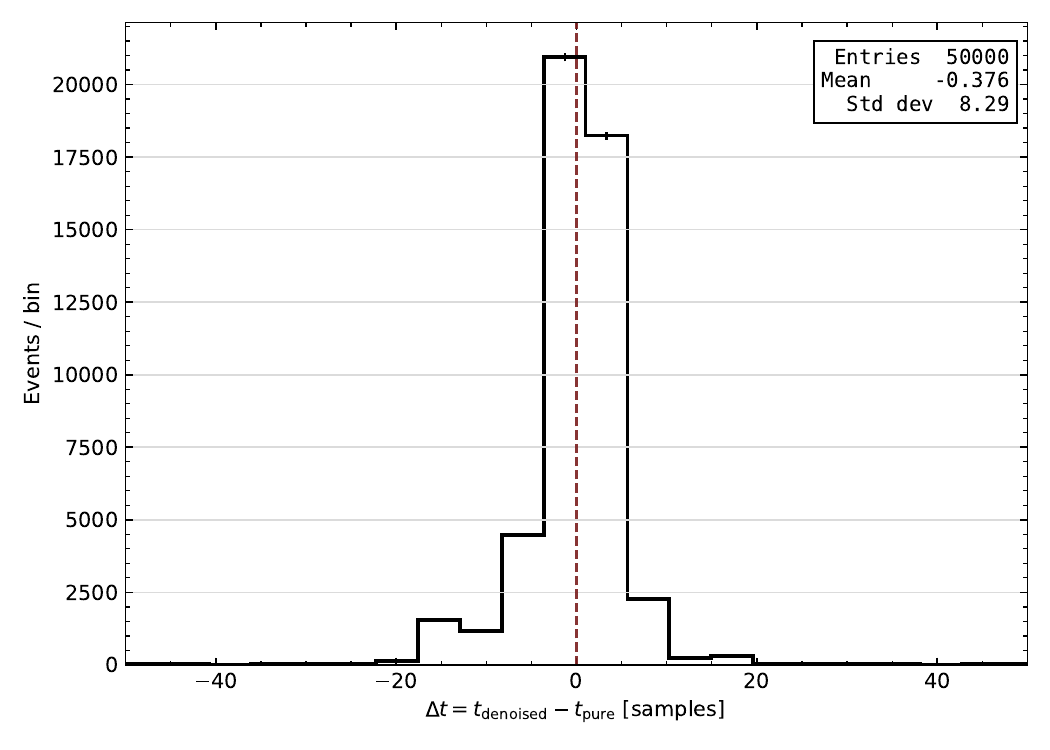}
    \caption{Pulse-time residual}
\end{subfigure}
\hfill
\begin{subfigure}{0.48\textwidth}
    \centering
    \includegraphics[width=\linewidth]{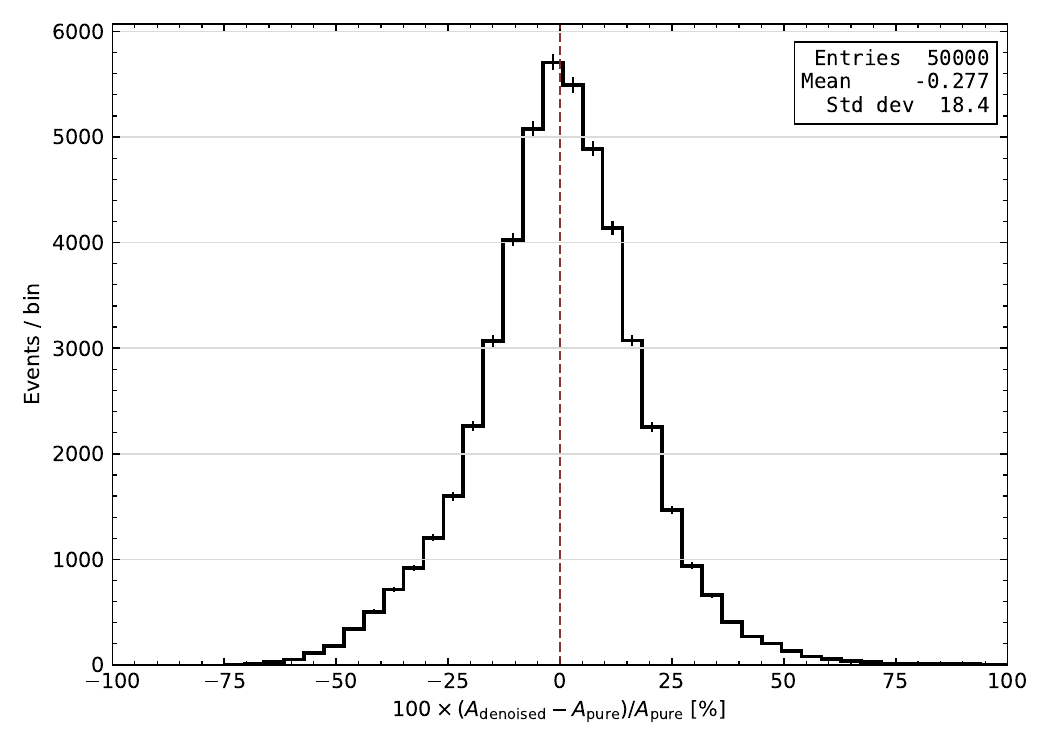}
    \caption{Peak-amplitude residual}
\end{subfigure}
\caption{Event-by-event feature residuals for denoised \texttt{signal} traces relative to the paired \texttt{pure signal} pulse. The timing residual tests whether the denoiser preserves the pulse position, while the amplitude residual tests whether the cleaned waveform keeps the pulse scale without relying on the downstream classifier score.}
\label{fig:denoiser_feature_diagnostics}
\end{figure}

Figure~\ref{fig:denoiser_feature_diagnostics} therefore tests the denoised waveform as an edge-processing product rather than only as an input to the classifier. The timing panel gives the residual between the envelope-peak position after denoising and the envelope-peak position of the paired pure pulse. Its concentration around zero supports the interpretation that the denoiser preserves the pulse location rather than merely suppressing large samples. The amplitude panel shows the corresponding peak-amplitude residual after denoising. The distribution is centred close to zero but has a finite width, which is the waveform-level origin of the residual spread summarised below in Table~\ref{tab:denoiser_feature_diagnostics}. These tails are important because fixed-threshold and feature-transmission schemes are usually controlled by the worst-behaved part of the distribution, not only by the mean response.

For the feature diagnostics, the denoised output is first mapped back to trace units using the z-score scale of the corresponding \texttt{signal} trace. The amplitude scale is then adjusted in post-analysis by one global factor, chosen from the median peak response of the denoised validation traces relative to their paired \texttt{pure signal} traces. This calibration is used only to display and quantify the denoised waveform in trace units; it is not used by the trigger decision or by the ROC curves. The timing residual is largely insensitive to such an overall amplitude rescaling, whereas the peak-amplitude and integrated-envelope diagnostics should be interpreted with this diagnostic calibration convention.

The same feature information can be compressed into the aggregate comparison in Table~\ref{tab:denoiser_feature_diagnostics}. Before denoising, the peak-amplitude response is biased high by the noise contribution, with a mean response factor of $1.92$ and a median response factor of $1.91$. After denoising, the mean peak-amplitude response is $0.997$ and the median is $1.00$. The pulse-time residual also improves, from a standard deviation of $41.3$ samples for the raw trace to $8.29$ samples after denoising, with the median residual reduced from one sample to zero. The integrated-envelope response remains above unity after denoising, with a mean value of $1.23$, so the denoised waveform should not be interpreted as a precision calorimetric reconstruction. For triggering and edge preprocessing, however, the relevant result is that the dominant timing and peak-amplitude distortions of the raw trace are substantially reduced.

\begin{table}[t]
\centering
\small
\caption{Aggregated waveform-level diagnostics for the denoiser on held-out \texttt{signal} traces. The raw noisy trace and the denoised trace are each compared to the paired \texttt{pure signal} target. Entries with scale ``samples'' are signed differences in pulse position, while entries with scale ``ratio'' are dimensionless response ratios relative to the paired target quantity. These quantities are diagnostic only and are not used to define the trigger operating thresholds.}
\label{tab:denoiser_feature_diagnostics}
\begin{tabular}{lccc}
\toprule
Diagnostic quantity & Raw signal vs.\ truth & Denoised vs.\ truth & Scale \\
\midrule
Pulse-time residual, mean & $2.18$ & $-0.38$ & samples \\
Pulse-time residual, std. dev. & $41.3$ & $8.29$ & samples \\
Pulse-time residual, median & $1.00$ & $0.00$ & samples \\
Peak-amplitude response, mean & $1.92$ & $0.997$ & ratio \\
Peak-amplitude response, std. dev. & $0.316$ & $0.184$ & ratio \\
Peak-amplitude response, median & $1.91$ & $1.00$ & ratio \\
Integrated-envelope response, mean & $6.25$ & $1.23$ & ratio \\
Integrated-envelope response, median & $5.07$ & $1.13$ & ratio \\
\bottomrule
\end{tabular}
\end{table}

\begin{figure}[t]
\centering
\begin{subfigure}{0.48\textwidth}
    \centering
    \includegraphics[width=\linewidth]{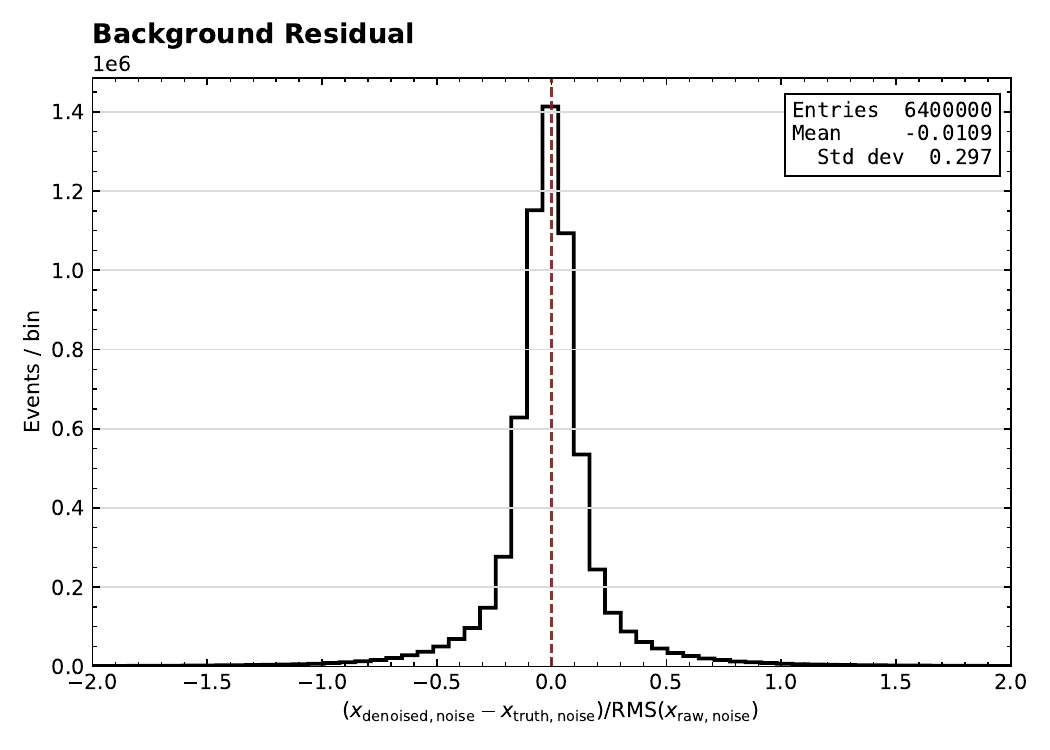}
    \caption{\texttt{background} traces}
\end{subfigure}
\hfill
\begin{subfigure}{0.48\textwidth}
    \centering
    \includegraphics[width=\linewidth]{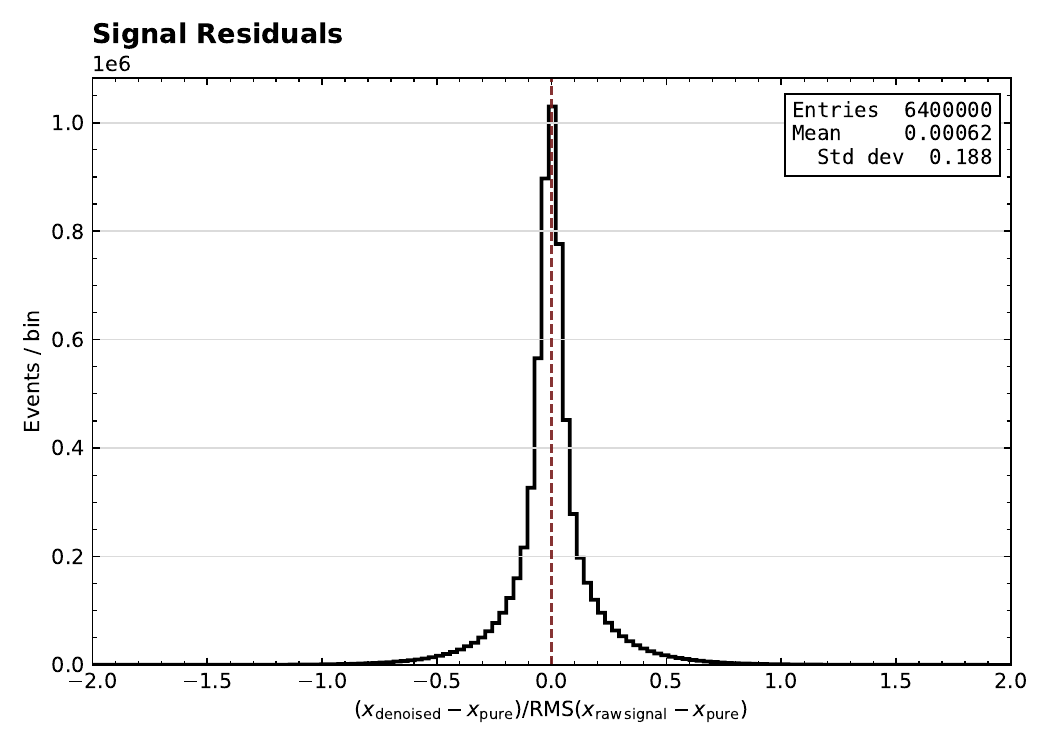}
    \caption{\texttt{signal} traces}
\end{subfigure}
\caption{Normalised sample-level denoiser residuals. For \texttt{background} traces the target is zero output and the residual is divided by the raw-noise RMS. For \texttt{signal} traces the residual is computed relative to the paired \texttt{pure signal} pulse and divided by the RMS of the injected noise contribution.}
\label{fig:denoiser_residual_diagnostics}
\end{figure}

The corresponding sample-level residual distributions are shown in figure~\ref{fig:denoiser_residual_diagnostics}. For each trace, the residual is normalised by the root-mean-square (RMS) of the relevant raw noise contribution: the raw noise trace for \texttt{background} samples and the difference between the raw \texttt{signal} trace and the paired \texttt{pure signal} pulse for signal-containing samples. The noise-only panel therefore tests the failure mode that is most relevant for a trigger, namely whether the denoiser creates signal-like structure from \texttt{background} traces. Over \num{6400000} samples, the residual distribution has a mean of $-0.0109$ and a standard deviation of $0.297$ in units of the raw-noise RMS. The signal panel shows the residual between the denoised waveform and the paired pure pulse, normalised in the same way to the injected noise contribution; its mean is $6.2\times10^{-4}$ and its standard deviation is $0.188$. The small biases and sub-noise RMS widths in both panels support the interpretation that the denoiser suppresses noise while preserving the pulse, rather than acting as an unconstrained pulse generator.

The practical gain is therefore twofold. First, denoising improves the trigger response when used before the classifier, increasing both the integrated AUC and the fixed-FPR efficiencies relative to the classifier-only branch. Second, the cleaned waveform is an intermediate data product. It can support simple peak-envelope decisions, timing diagnostics, event-quality checks, or selective waveform readout, whereas a decision-level trigger only returns a score. This distinction is important for bandwidth-constrained autonomous radio stations: the denoiser is not only another way to make the same binary decision, but a compact front-end transformation that makes weak pulses more accessible to both the trigger and the downstream readout logic.

\subsection{Limitations and robustness checks}
\label{sec:limitations}

The implemented chain demonstrates that the denoiser--classifier trigger can run with the normalised trace representation required by the HGQ-based quantised models used here. For deployment, this normalisation should be treated as part of the trigger interface: the station trigger must either receive or compute the same representation used during model validation. For a compact quantised FPGA trigger, the amplitude range and normalisation rule define the numerical range seen by the neural models. The deployment version can therefore specify this interface together with the firmware, using for example station-level pedestal and scale constants, a cheaper fixed calibration scheme, or a retrained/exported model variant operating on calibrated ADC-scale inputs. This normalisation interface is also a near-term hardware optimisation opportunity. The resource numbers reported for the implemented full chain include the normalisation stage, so the quoted DSP count should be interpreted as the cost of the realised trigger path rather than of the denoiser and classifier convolutional core alone. Reducing the arithmetic cost of the normalisation stage would lower the full-chain DSP demand without changing the basic denoiser--classifier architecture.

The same interface matters if the denoised waveform is used beyond the station-level trigger decision. The timing diagnostics in Sec.~\ref{sec:why_denoiser} test whether denoising preserves the pulse position, which is the waveform feature most relevant for possible array-level, polarisation-level, or multi-channel coincidence logic. For peak-amplitude and integrated-envelope features, a deployment implementation should either remove the normalisation step by exporting a model that operates on calibrated ADC-scale inputs, or define the inverse scale from station calibration and background-monitoring quantities available in the firmware. This would replace the present analysis-side amplitude calibration with an operational calibration path and allow the waveform-level validation to be repeated directly in deployment units for feature-only readout, selective waveform buffering, and other bandwidth-saving modes.

The signal benchmark is intentionally weighted toward weak traces. This choice provides a stress test of the operating regime in which a radio-only trigger is most constrained. The earlier classifier-only trigger already performs well for high-SNR pulses, whereas the relevant question for the present work is whether denoising recovers efficiency close to threshold at fixed background rate. The main trigger conclusion is therefore taken from the fixed-FPR efficiency comparison against the same threshold baseline, the improved integrated ROC performance, and the dependence of this gain on signal strength.

We checked the stability of this conclusion with two alternative signal mixtures. A validation sample shifted toward higher signal strengths, spanning approximately \(-0.8 \lesssim \log_{10}(\mathrm{SNR}) \lesssim 2.7\), gives the expected behaviour: the raw peak-envelope threshold trigger improves because the sample contains fewer near-threshold traces, while the hybrid branch remains more efficient, with an absolute TPR increase of about 33 percentage points at FPR \(=10^{-4}\). In a second check, the event-by-event amplitude scan was replaced by a single global scale factor applied to the detector-level pulse library, leaving the natural relative pulse-amplitude variation of the simulated traces unchanged. This sample has a different signal-strength composition and is therefore not expected to reproduce the benchmark efficiency curve bin by bin. The relevant observation is instead that the trigger ordering is preserved. In common signal-strength bins with sufficient statistics, the hybrid efficiency remains several times larger than the raw peak-trigger efficiency; representative overlapping bins at the nominal FPR of \(10^{-2}\) give hybrid efficiencies of about \(0.68{-}0.89\), compared with raw peak-trigger efficiencies of about \(0.07{-}0.10\), corresponding to gains of order \(7{-}9\). These checks support the interpretation that the observed hybrid-trigger advantage is not an artefact of the particular near-threshold amplitude weighting used for the main benchmark, while the absolute efficiency values remain tied to the stated benchmark definition.

As an additional detector-environment check, the benchmark was repeated using measured AERA station background with the same signal-injection and SNR-construction procedure. The trigger performance was moderately improved under the investigated conditions. Since the signal amplitudes were populated according to the same SNR construction, the lower absolute AERA noise level does not by itself produce an orders-of-magnitude improvement; the observed improvement is consistent with the smaller contribution of impulsive, pulse-like disturbances in the AERA background.

The background validation should also be extended toward detector generalisation. The measured-noise sample used here is valuable because it provides a controlled comparison to the earlier classifier-only trigger under a deliberately difficult interference environment. In a deployed radio array, however, stations may operate under heterogeneous noise conditions due to site-dependent RFI, antenna and electronics differences, seasonal variations, and time-dependent interference. A natural next study is therefore to test the same fixed-FPR protocol across multiple stations and data-taking periods, and to compare a single globally trained trigger with deployment strategies that include station-level calibration constants, periodic retraining, domain adaptation, or controlled online updates based on background data streams. This would address the practical question of how much model adaptation is needed to preserve the demonstrated low-FPR trigger gain across a heterogeneous detector. For an air-shower rate measurement, noise conditions---including the RMS level and the occurrence of impulsive interference---must be monitored, and the signal efficiency calibrated by station and data-taking period, so that changes in trigger acceptance are not mistaken for changes in the physical rate.

Finally, the present trigger is deliberately formulated as a single-channel, fixed-window edge block. This is the right level for isolating the effect of denoising on the waveform and for producing a compact FPGA implementation. The reported hardware results therefore characterise the processing of an already formed \num{128}-sample window; continuous-stream window formation, buffering, and scheduling are system-level functions not included in the quoted resource and power values. A station or array deployment can add further information, including multi-antenna coincidence, polarisation, station geometry, and timing consistency across channels. The denoised waveform is useful in that extension because it can support not only a station-level decision, but also feature-only readout, selective waveform buffering, and compact feature transmission to an array-level trigger in a streaming setup with realistic buffering, threshold updates, deadtime, and data-acquisition constraints. Translating the resulting station-level gain into detector exposure or event rate is then a separate array-level study.

%% file: sections/07_summary_and_outlook.tex
\section{Summary and outlook}
\label{sec:conclusion}

This paper describes a hybrid neural trigger developed for autonomous radio detection of extensive air showers in a high-interference environment. The central idea is to move the trigger from a pure decision problem to a waveform-recovery problem followed by a decision: a compact denoiser first suppresses the measured radio background and recovers the air-shower pulse candidate, and a compact classifier then evaluates the cleaned trace. The trigger is studied on a deliberately difficult benchmark built from measured noise and detector-folded simulated pulses from the Pierre Auger \emph{Offline} chain, with the signal population concentrated in the near- and sub-threshold regime where autonomous radio triggering is most constrained.

The resulting comparison shows why the denoiser is not merely an auxiliary preprocessing layer. When the same peak-envelope decision is applied after denoising, the area under the receiver-operating-characteristic curve rises from \(0.63\) for the raw trace to \(0.98\) for the denoised trace. This control test isolates the waveform-cleaning stage from the classifier and shows that the denoiser changes the trace representation in a trigger-relevant way. In the full denoiser--classifier chain, the hybrid trigger improves the overall ROC performance and the fixed-FPR efficiencies: at a false-positive rate of \(10^{-4}\), it retains about \(41\%\) of the held-out signal traces in the weak-signal benchmark, compared with \(27\%\) for the classifier acting directly on the raw trace and about \(2\%\) for the peak-envelope threshold reference.

The cleaned waveform is also a useful station-level data product. Noise-only validation shows that the denoiser suppresses background traces without producing signal-like pulse structure, while signal-trace diagnostics show that the pulse time and peak-amplitude response are substantially stabilised relative to the raw waveform. The timing residual width decreases from \(41.3\) samples before denoising to \(8.29\) samples after denoising, and the median peak-amplitude response after denoising is consistent with unity under the current analysis-side normalisation. This does not make the denoiser a precision energy-reconstruction tool, but it does make the cleaned trace useful for timing checks, quality flags, feature-only readout, selective waveform transmission, and downstream coincidence logic.

A second part of the contribution is the deployment workflow. The denoiser and classifier are not selected only for floating-point performance and then compressed afterwards. Hyperparameter optimisation, model-size constraints, quantisation-aware training, high-granularity fixed-point quantisation, \texttt{hls4ml} export, high-level synthesis, and register-transfer-level validation are used as a connected development chain. The implemented firmware preserves the software trigger response and reaches placed-and-routed implementations on representative field-programmable gate array targets. The constrained reference implementation meets the timing target with microsecond-scale latency and uses only tens of dedicated arithmetic blocks, demonstrating that the hybrid trigger is not only a physics-motivated algorithm but also a realistic edge-hardware design.

The main limitations define the next steps. The trace normalisation and inverse-normalisation convention must be fixed as part of the firmware interface, preferably through station-level calibration constants or a model trained directly in calibrated trace units. The background validation should be extended across sites, antennas, electronics chains, seasons, and time-dependent interference conditions to determine how much retraining or domain adaptation is needed. Finally, the single-channel fixed-window trigger should be embedded in a station- and array-level architecture that includes polarisation, multi-antenna timing consistency, buffering, deadtime, threshold updates, and data-acquisition constraints.

Within the scope of this study, the conclusion is that hybrid neural denoising provides both a more sensitive weak-pulse trigger representation and a deployable field-programmable gate array workflow. This makes it a concrete route toward radio-only self-triggering for inclined and otherwise radio-faint air showers in environments where conventional threshold triggers are limited by radio background. 

%% file: sections/08_acknowledgments.tex
The authors gratefully acknowledge the support of the Electronics Laboratory of the Department Physik at Universit\"at Siegen for assistance with the experimental electronics, laboratory infrastructure, and hardware-oriented validation work that underpins the measured-noise and FPGA studies. We thank Prof. Katie Mulrey and Prof. Harm Schoorlemmer for providing the antenna used in the measurement campaign, and Noah Siegemund for assistance with the hardware setup on campus.

We acknowledge the Pierre Auger Collaboration for the simulation and detector-response tools on which the detector-aware signal construction is based, and for the air-shower simulation framework used to produce the detector-folded pulse library. We thank the \texttt{hls4ml} team, and in particular Enrico Lupi, for helpful discussions on the use and configuration of \texttt{hls4ml}. We also acknowledge the developers and maintainers of the open-source software used throughout this work, including \texttt{Python}, \texttt{NumPy}, \texttt{SciPy}, \texttt{Matplotlib}, \texttt{TensorFlow}/\texttt{Keras}, \texttt{QKeras}, \texttt{hls4ml}, \texttt{cocotb}, \texttt{Vitis HLS}, and \texttt{Vivado}.